\documentclass[fleqn,usenatbib]{mnras}

\usepackage{newtxtext,newtxmath}

\usepackage[T1]{fontenc}

\defcitealias{Dhabal2019}{D19}

\usepackage{lipsum}  

\DeclareRobustCommand{\VAN}[3]{#2}
\let\VANthebibliography\thebibliography
\def\thebibliography{\DeclareRobustCommand{\VAN}[3]{##3}\VANthebibliography}

\usepackage{graphicx}	
\usepackage{amsmath}	

\title[Relative Alignments in NGC 1333]{Relative Alignments Between Magnetic Fields, Velocity Gradients, and Dust Emission Gradients in NGC 1333}

\author[M. Chen et al.]{
Michael Chun-Yuan Chen,$^{1}$\thanks{E-mail:chen.m@queensu.ca}
Laura M. Fissel,$^{1}$
Sarah I. Sadavoy,$^{1}$
Erik Rosolowsky,$^{2}$
Yasuo Doi,$^{3}$
\newauthor
\ Doris Arzoumanian,$^{4}$
Pierre Bastien,$^{5}$
Simon Coud\'{e},$^{6,7}$
James Di Francesco,$^{8,9}$
Rachel Friesen,$^{10}$
\newauthor
\ Ray S. Furuya,$^{11}$
Jihye Hwang,$^{12,13}$
Shu-ichiro Inutsuka,$^{14}$
Doug Johnstone,$^{8,9}$
Janik Karoly,$^{15}$
\newauthor
\ Jungmi Kwon,$^{16}$
Woojin Kwon,$^{17,18}$
Valentin J. M. Le Gouellec,$^{19,20}$
Hong-Li Liu,$^{21}$
Steve Mairs,$^{6}$
\newauthor
\ Takashi Onaka,$^{16}$
Kate Pattle,$^{22}$
Mark G. Rawlings,$^{23,24}$
Mehrnoosh Tahani,$^{25}$
Motohide Tamura,$^{4,16,26}$
\newauthor
\ Jia-Wei Wang,$^{27}$
and the BISTRO Team
\\
$^{1}$Department of Physics, Engineering Physics and Astronomy, Queen's University, Kingston, ON K7L 3N6, Canada\\
$^{2}$Department of Physics, University of Alberta, Edmonton, AB, T6G 2E1, Canada\\
$^{3}$Department of Earth Science and Astronomy, The University of Tokyo, 3-8-1 Komaba, Meguro, Tokyo 153-8902, Japan\\
$^{4}$National Astronomical Observatory of Japan, Osawa 2-21-1, Mitaka, Tokyo 181-8588, Japan \\
$^{5}$Institut de Recherche sur les Exoplan\`etes (iREx) \& Centre de Recherche en Astrophysique du Qu\'ebec (CRAQ),\\ \ Universit\'e de Montr\'eal, D\'epartement de Physique, 1375, Avenue Th\'er\`ese-Lavoie-Roux, Montr\'eal, QC, H2V 0B3, Canada \\
$^{6}$Department of Earth, Environment, and Physics, Worcester State University, Worcester, MA 01602, USA \\
$^{7}$Center for Astrophysics $\vert$ Harvard \& Smithsonian, 60 Garden Street, Cambridge, MA 02138, USA \\
$^{8}$NRC Herzberg Astronomy and Astrophysics, 5071 West Saanich Rd, Victoria, BC, V9E 2E7, Canada\\
$^{9}$Department of Physics and Astronomy, University of Victoria, Victoria, BC, V8P 5C2, Canada\\
$^{10}$David A. Dunlap Department of Astronomy and Astrophysics, University of Toronto, 50 St. George Street, Toronto, ON M5S 3H4, Canada\\
$^{11}$Institute of Liberal Arts and Sciences Tokushima University, Minami Jousanajima-machi 1-1, Tokushima 770-8502, Japan \\
$^{12}$Korea Astronomy and Space Science Institute, 776 Daedeokdae-ro, Yuseong-gu, Daejeon 34055, Republic of Korea\\
$^{13}$University of Science and Technology, Korea, 217 Gajeong-ro, Yuseong-gu, Daejeon 34113, Republic of Korea\\
$^{14}$Department of Physics, Graduate School of Science, Nagoya University, Furo-cho, Chikusa-ku, Nagoya, 464-8602, Japan \\
$^{15}$Jeremiah Horrocks Institute, University of Central Lancashire, Preston, PR1 2HE, UK \\
$^{16}$Department of Astronomy, Graduate School of Science, The University of Tokyo, 7-3-1 Hongo, Bunkyo-ku, Tokyo, 113-0033, Japan \\
$^{17}$Department of Earth Science Education, Seoul National University, 1 Gwanak-ro, Gwanak-gu, Seoul 08826, Republic of Korea \\
$^{18}$SNU Astronomy Research Center, Seoul National University, 1 Gwanak-ro, Gwanak-gu, Seoul 08826, Republic of Korea \\
$^{19}$NASA Ames Research Center, Space Science and Astrobiology Division M.S. 245-6 Moffett Field, CA 94035, USA\\
$^{20}$NASA Postdoctoral Program Fellow \\
$^{21}$Department of Astronomy, Yunnan University, Kunming, 650091, People's Republic of China \\
$^{22}$Department of Physics and Astronomy, University College London, London, WC1E 6BT, UK \\
$^{23}$East Asian Observatory, 660 N. A'oh\={o}k\={u} Place, University Park, Hilo, HI, 96720, USA \\
$^{24}$Gemini Observatory/NSF's NOIRLab, 670 N. A'oh\={o}k\={u} Place, Hilo, HI, 96720, USA \\
$^{25}$Kavli Institute for Particle Astrophysics \& Cosmology (KIPAC), Stanford University, Stanford, CA 94305, USA \\
$^{26}$Astrobiology Center, National Institutes of Natural Sciences, 2-21-1 Osawa, Mitaka, Tokyo, 181-8588, Japan \\
$^{27}$Academia Sinica Institute of Astronomy and Astrophysics, No.1, Sec.4, Roosevelt Road, Taipei 10617, Taiwan
}

\date{Accepted XXX. Received YYY; in original form ZZZ}

\pubyear{2023}

\begin{document}
\label{firstpage}
\pagerange{\pageref{firstpage}--\pageref{lastpage}}
\maketitle

\begin{abstract}
Magnetic fields play an important role in shaping and regulating star formation in molecular clouds. Here, we present one of the first studies examining the relative orientations between magnetic ($B$) fields and the dust emission, gas column density, and velocity centroid gradients on the 0.02 pc (core) scales, using the BISTRO and VLA+GBT observations of the NGC 1333 star-forming clump. We quantified these relative orientations using the Project Rayleigh Statistic (PRS) and found preferential global parallel alignment between the $B$ field and dust emission gradients, consistent with large-scale studies with \emph{Planck}. No preferential global alignments, however, are found between the $B$ field and velocity gradients. Local PRS calculated for subregions defined by either dust emission or velocity coherence further revealed that the $B$ field does not preferentially align with dust emission gradients in most emission-defined subregions, except in the warmest ones. The velocity-coherent structures, on the other hand, also showed no preferred $B$ field alignments with velocity gradients, except for one potentially bubble-compressed region. Interestingly, the velocity gradient magnitude in NGC 1333 ubiquitously features prominent ripple-like structures that are indicative of magnetohydrodynamic (MHD) waves. Finally, we found $B$ field alignments with the emission gradients to correlate with dust temperature and anticorrelate with column density, velocity dispersion, and velocity gradient magnitude. The latter two anticorrelations suggest that alignments between gas structures and $B$ fields can be perturbed by physical processes that elevate velocity dispersion and velocity gradients, such as infall, accretions, and MHD waves.

\end{abstract}




\section{Introduction}

Over the past decades, evidence has emerged that filaments play a crucial role in how molecular clouds form stars (see \citealt{Andre2014}; \citealt{Pineda2023}). Analytically, filaments have the most favourable geometry for the growth of local internal structure before becoming overwhelmed by global collapse \citep{Pon2011}. Numerically, filaments are produced or seeded naturally by supersonic turbulence within molecular clouds \citep[e.g.,][]{Porter1994,Vazquez-Semadeni1994}, after which they fragment into star-forming cores \citep[e.g.,][]{Seifried2015}. In the absence of turbulence strong enough to support the cloud globally, the cloud can also collapse hierarchically, channelling accretion flows continuously via filaments from the larger scales to cores and proto-cluster hubs \citep{Vazquez-Semadeni2019}. Indeed, star-forming structures have been observed to be hierarchically filamentary \citep{Hacar2023} and velocity gradients observed across and along filaments have indicated ongoing mass flows onto and along filaments, respectively \citep[e.g.,][]{Peretto2006, Kirk2013, Friesen2013, Shimajiri2019, ChenMike2020}.

The assembly of star-forming structures, however, is not governed by gravity and turbulence alone. Numerous theoretical studies on magnetohydrodynamics (MHD) have shown that magnetic fields can also play a substantial role alongside gravity \citep[e.g.,][]{Nagai1998,Nakamura2008,Soler2013,Inoue2018,Pattle2023}, and from dust polarization observations, dense filamentary structures tend to align perpendicularly with the magnetic fields \citep[e.g.,][]{Palmeirim2013, Planck2016_HROMagClouds, Fissel2019, Soam2019, Pattle2023}. These results conform to the expectations of idealized MHD, where gas can flow freely along magnetic field lines but not against them, which causes gas to accumulate into filamentary structures that mainly align perpendicularly to the magnetic field \citep[e.g.,][]{Inutsuka2015, Gomez2018}.

Molecular line observations have also revealed velocity gradients along large-scale magnetic fields indicating magnetically-mediated accretion flows \citep[e.g.,][]{Palmeirim2013, Bonne2020}. Statistical comparisons between the local orientations of the magnetic field and velocity gradients \citep[e.g.,][]{TangYW2019, WangJW2020}, however, are very rare and lack the resolution to resolve the most common filaments found by \textit{Herschel} studies \citep[$\sim 0.1$ pc in width; e.g.,][]{Arzoumanian2011, Arzoumanian2019}. Considering that magnetic fields can be significantly distorted by gas structures contracting under self-gravity and consequently alter the direction of gas flow \citep{Gomez2018}, observations that can investigate the relationship between gas flows, gas morphology, and the magnetic field on the scales of both filaments and cores are crucial. Such studies will not only test how magnetic fields guide gas flows (e.g., \citealt{Seifried2015}) to form and feed star-forming structures (e.g., \citealt{Gomez2018}) but also test how they provide support against gravity (e.g., \citealt{Nakano1978}) or mediate protostellar feedback (e.g., \citealt{Offner2018}).

In this paper, we investigate the role of the magnetic field in the active star-forming complex NGC 1333 using combined dust polarization, dust emission, H$_2$ column density, and NH$_3$ line emission observations on $\sim 0.02$ pc scales. Specifically, we make statistical orientation comparisons between the magnetic field and the velocity, emission, and column density gradients. Located at $299 \pm 17$ pc away \citep{Zucker2018} in the Perseus Molecular Cloud, NGC 1333 is one of the most extensively studied cluster-forming clumps in the nearby neighbourhood \citep{Walawender2008}, with very well-characterized populations of dense cores \citep[e.g.,][]{Pezzuto2021} and young stellar objects \citep[YSOs; e.g.,][]{Dunham2015}. The star-forming clump consists of a network of supercritical filaments \citep{Hacar2017} that show velocity gradients indicative of accretion onto and along the filaments \citep{ChenMike2020}. The clump appears to be collapsing globally \citep{Walsh2006}, which likely explains why NGC 1333 has a higher velocity dispersion between its filaments than those found in other Perseus clumps (Chen et al. 2023, submitted). Recently, the magnetic field in NGC 1333 has also been studied on $\sim 0.02$ pc scales, revealing a highly ordered plane-of-the-sky (POS) magnetic field with respect to the clump's filaments \citep{Doi2020}. 

The paper is structured as follows: In Section \ref{sec:obs}, we describe the observations and data reduction of our NH$_3$ (1,1) and dust polarization data, as well as the dust emission and column density maps obtained from publicly available archives. Section \ref{sec:methods} describes our multi-component fits to the NH$_3$ data and explains how we identify velocity-coherent structures from these fits and compute gradients. In Section \ref{sec:methods}, we also outline the statistical methods we use to quantify the relative orientations (ROs) between the POS magnetic field (hereafter, the magnetic field unless stated otherwise) and our computed gradient fields. We present our results and discussions in Section \ref{sec:results} and \ref{sec:discussion}, respectively, and summarize our conclusions in Section \ref{sec:conclusions}.
 
\section{Observations \& Data Reduction}
\label{sec:obs}

\subsection{Ammonia data}
\label{subsec:nh3_data}

We use spectral line observations of ammonia (NH$_3$) from the Karl G. Jansky Very Large Array (VLA) project 13A-309 (PI: Shaye Storm), which surveyed the NGC 1333 region using the VLA D-configuration in seven observational tracks.  The observations are discussed in detail in \citet[][hereafter, \citetalias{Dhabal2019}]{Dhabal2019}. In this work, we follow a similar reduction and imaging strategy.  We first calibrate the measurement sets using the VLA pipeline released with CASA version 4.7.2 \citep{casa}, yielding gain calibrated and flagged visibilities ready for imaging.  We imaged the data with CASA version 5.1.0-29 using the \textsc{tclean} algorithm, setting a spectral channel width of 6.9 kHz ($\sim 0.085$ km s$^{-1}$). The imaging used a multiscale deconvolution algorithm with scales of $0$ and  $0.66'' \times 2^{k}$ for $k=0\cdots 5$ with a threshold of 15 mJy/beam and \citet{briggs} weighting using \texttt{robust}$=0.5$.

To improve the quality of the imaging process, we create a clean mask to restrict the regions where the deconvolution can identify clean components.  For the clean mask, we use single-dish data of the region in NH$_3$ from the Robert F. Byrd Green Bank Telescope (GBT) that were collected as part of the Green Bank Ammonia Survey \citep[GAS,][]{GAS}. We define the clean mask using regions of significant emission ($T_\mathrm{MB}>5 \sigma_T$) in the GBT data, and then we deconvolve the VLA data using the clean mask, cleaning each channel down to 15 mJy ($\sim 2.5\sigma$).  The restoring beam is $3.3''\times 2.8''$ FWHM.

Following deconvolution, we use the CASA task \textsc{feather} to combine the VLA and GBT data in the Fourier domain, yielding a fully sampled image for analysis. The \textsc{feather} task assumes that the flux calibration between the two telescopes is accurate. We used the diagnostics in the \textsc{uvcombine} package \citep{uvcombine} to check this assumption, finding the ratio of the flux scales to be consistent with unity. \citetalias{Dhabal2019} use a different approach to integrating information from the GBT, combining the data in the visibility domain before deconvolution. This difference in method is not expected to lead to substantial ($>10\%$) variation in the final image products \citep{SDcomb}.

In this work, we focus on the NH$_3$(1,1) line at $\nu_0=23.694$~GHz though we carried out imaging for NH$_3$(2,2) and NH$_3$(3,3) as well. The final VLA+GBT maps also have a resolution of $3.3''\times 2.8''$ and a noise level of 6.5 mJy~beam$^{-1}$, which is equivalent to 1.5~K at this resolution. To match the spatial resolution of our polarization data, we further convolved our NH$_3$(1,1) to a resolution of $14.1''$ using a Gaussian kernel and re-gridded the convolved image with 2.7\arcsec pixels.

\subsection{Polarization data}
\label{subsec:pol_data}

Our 850 \textmu m dust polarization measurements of NGC 1333, i.e., Stokes \textit{I}, \textit{Q}, and \textit{U} and their estimated uncertainties, were first published by \cite{Doi2020} as part of the B-Fields in Star Forming Regions (BISTRO) Survey \citep{Ward-Thompson2017} using the POL-2 instrument on the James Clerk Maxwell Telescope (JCMT). The effective beam size of the observations is $14.1''$, with the final pixel size of the data product being $7.05''$. Following the quality control set by \cite{Doi2020}, we estimated the polarized intensity (\textit{PI}) and its error ($\delta PI$) as follows:
\begin{equation}\label{eq:PI}
PI = \sqrt{Q^2 + U^2},
\end{equation}
\begin{equation}\label{eq:PI_error}
\delta PI = \frac{ \sqrt{(Q^2 \cdot \delta Q^2 + U^2 \cdot \delta U^2)}}{PI},
\end{equation}
where $\delta Q$ and $\delta U$ are the estimated errors of $Q$ and $I$, respectively. While the $\delta PI$ estimated above is biased by the squared $Q$ and $U$ offsetting the derived $PI$, such a bias is negligible when the SNR is high (e.g. $PI / \delta PI > 3$; \citealt{Wardle1974}; \citealt{Naghizadeh-Khouei1993}; \citealt{Vaillancourt2006}). For this reason and quality assurance, only pixels with $PI / \delta PI > 3$ and $I > 25$ mJy beam$^{-1}$ (i.e., $I / \delta I > 10$) are used for our analysis. We note that the polarization angles, which are the primary focus of our analyses, should not be affected by such bias either way. The typical rms noise levels for $I$, $Q$, and $U$ in our samples are 1.1 mJy beam$^{-1}$, 0.9 mJy beam$^{-1}$, and 0.9 mJy beam$^{-1}$, respectively.

For this analysis, we primarily focus on the magnetic field orientation. Following \cite{Doi2020}, we measure polarization angles ($\psi$) and their associated errors ($\delta \psi$) as:
\begin{equation}\label{eq:PA}
\psi = \frac{1}{2}\arctan{\left(\frac{U}{Q}\right)},
\end{equation}
\begin{equation}\label{eq:PA_error}
\delta \psi = \frac{1}{2}\frac{ \sqrt{(Q \cdot \delta U)^2 + (U \cdot \delta Q)^2}}{PI^2}.
\end{equation}
With our imposed SNR thresholds for the polarized data, the polarization angle uncertainties are $0.2^\circ < \delta \psi < 9.6^\circ$ for all pixels, with a median and mean uncertainty of $\delta\psi \approx 6^\circ$.

Since thermal dust emission is expected to be polarized relative to the long-axis of the dust grains, with dust grains' short axes aligned with the magnetic field \citep[e.g.,][]{Andersson2015}, we rotate all polarization angles by 90$^\circ$ to obtain the inferred magnetic field orientation. We designate the unit half-vector $\mathbf{\hat{B}}$ to represent magnetic field orientation measurements (hereafter, $B$ field measurements).

\subsection{Dust emission, Column Density, \& Dust Temperature Data}
\label{subsec:herschel_data}

Since the Stokes \textit{I} map released by \cite{Doi2020} does not recover extended emission well due to atmospheric filtering and a slow scan rate \citep{Chapin2013, Friberg2016}, we adopted the JCMT Gould Belt Survey (JGBS) Data Release 3 (DR3) intensity map \citep{KirkH2018} for our structural analysis of the 850 \textmu m dust emission. The DR3 data product has a median rms of $\sim 12.1$ mJy beam$^{-1}$ and a pixel size of $3''$. While the JGBS 850 \textmu m map is less sensitive than the BISTRO total intensity map, it has less aggressive spatial filtering and these data have been corrected for line contamination from $^{12}$CO ($2 - 1$) using methods laid out by \cite{Drabek2012} and \cite{Parsons2018}. Since CO contamination can be substantial in NGC 1333 (see Appendix \ref{appendix:CO}), the JGBS data product is more appropriate for performing our structural analysis. Considering that we do not expect CO to be polarized, any $^{12}$CO ($2 - 1$) contribution should not result in spurious polarization in the BISTRO polarization maps.

We obtained the \textit{Herschel} 160 \textmu m and 250 \textmu m dust emission maps, as well as the \textit{Herschel}-derived dust temperature ($T_\mathrm{d}$), and high-resolution H$_2$ column density, $N($H$_2$), maps of Perseus from the \textit{Herschel} Gould Belt Survey (HGBS) Archive. These archival data were first published by \cite{Pezzuto2021}. The resolutions of the 160 \textmu m and 250 \textmu m dust emission maps, as well as the \textit{Herschel}-derived $T_\mathrm{d}$ and high-resolution $N(\mathrm{H}_2)$ maps, are $13.5''$, $18.2''$, $36.3''$, and $18.2''$, respectively.

\section{Methods}
\label{sec:methods}

\subsection{Multi-component line fitting}

To derive gas kinematic properties, we fit our NH$_3$ (1,1) data with spectral models containing up to two velocity components using the \texttt{MUFASA} package \citep{ChenMike2020}. \texttt{MUFASA} is an automated iterative fitter that wraps around the \texttt{PySpecKit} package \citep{Ginsburg2022} to perform least-squares fits using the Levenberg–Marquardt method (\citealt{Levenberg1944}; \citealt{Marquardt1963}; \citealt{More1978}). \texttt{MUFASA} initially convolves the cube to half its original angular resolution (i.e., doubles the beam size) and then fits the convolved cube using initial guesses produced by a moment-based technique detailed in \cite{ChenMike2020}. The resulting fits from the convolved cube are then adopted as the updated initial guesses for the next iteration of fits. This two-iteration approach uses the convolved cube's boosted signal and enhanced spatial awareness to improve the fits. We adopt the same \texttt{MUFASA} NH$_3$ (1,1) model as \citet{ChenMike2020, ChenMike2022} that treats each velocity component as a slab of gas under local thermodynamic equilibrium (LTE) and features a Gaussian velocity profile. Each velocity component of the model is parameterised by its velocity centroid ($v_\mathrm{LSR}$), velocity dispersion ($\sigma_ \mathrm{v}$), excitation temperature ($T_\mathrm{ex}$), and total optical depth ($\tau_{0}$). The model also accounts for all 18 hyperfine components of the NH$_3$ (1,1) transition for each velocity component and performs radiative transfer through each component along the line of sight from the cosmic microwave background towards the observer. 

To determine whether a spectrum is robustly detected via statistical model selection, \texttt{MUFASA} further assumes an emission-free noise model of a flat spectral baseline centred on a specific intensity of zero for comparison. \texttt{MUFASA} selects the best fit between noise, one-component, and two-component models using the corrected Akaike Information Criterion (AICc; \citealt{Akaike1974}; \citealt{Sugiura1978}). The software selects model \textit{b} over model \textit{a} when their log relative likelihood $K$ obtained from the AICc, expressed as 
\begin{equation}\label{eq:lnk}
\ln{K_a^b} = - \left ( \textup{AICc}_b - \textup{AICc}_a \right )/2,
\end{equation}
is above a statistically robust threshold of 5 \citep{Burnham2004}. \texttt{MUFASA} calculates the AICc values from the residual sum of squares ($\mathrm{RSS}$) of the fit as
\begin{equation}\label{eq:AICc}
\mathrm{AICc} = n\ln{\left(\frac{\mathrm{RSS}}{n}\right) + 2p + \frac{2p(p+1)}{n-p-1}},
\end{equation}
where $n$ and $p$ are the sample size of the spectra and the number of model parameters, respectively. \texttt{MUFASA}'s model selection with AICc performs comparably to the full Bayesian approach on the same NH$_3$ data \citep[see][]{Sokolov2020, ChenMike2020}. The AICc approach, however, is more efficient because it does not need to sample the likelihood space exhaustively like the Bayesian approach.

\subsection{Identifying velocity-coherent structures}\label{subsec:vcs_def}

Star-forming gas can often contain multiple velocity-coherent structures, hereafter VCS, which are identified as gas regions with smoothly varying kinematic properties. These kinematically distinct structures have been found in observations with tracers such as CO \citep[e.g.,][]{Hacar2013} and NH$_3$ \citep{ChenMike2020}. To identify these structures, we employ the Density-Based Spatial Clustering of Applications with Noise algorithm \citep[\texttt{DBSCAN};][]{Ester1996} on the \texttt{MUFASA}-derived $v_\mathrm{LSR}$ to identify clusters in Position-Position-Velocity (PPV) space \citep[e.g., similar to][]{Doi2020}. Specifically, we use \texttt{DBSCAN} as implemented by the \texttt{scikit-learn} package \citep{scikit-learn}. Clustering algorithms like \texttt{DBSCAN} group data points (i.e., samples) in N-dimensional parameter space based on their proximity to one another in that space. When applied to samples in PPV space, \texttt{DBSCAN} will identify spatially and kinematically continuous structures.

\texttt{DBSCAN}, compared to many clustering algorithms, has the advantage of being robust at identifying clusters with irregular shapes, not requiring a user-specified number of clusters, and not forcing a cluster membership onto all data samples. This last attribute allows \texttt{DBSCAN} to identify ``noise'' (i.e., outlier) members in addition to cluster members. Moreover, \texttt{DBSCAN} primarily operates only on two user parameters: 1) $\varepsilon$, the maximum distance threshold to qualify two samples as being in the same neighbourhood and 2) $s_\mathrm{min}$, the minimum samples in a neighbourhood required to consider these samples core points (i.e., a seed). A cluster can only be seeded if it contains at least $s_\mathrm{min}$ number of samples within a $\varepsilon$ distance of each other (i.e., core points) and grows subsequently to include all samples directly reachable via a core point within a distance of $\varepsilon$. With so few parameters, \texttt{DBSCAN} is an attractive clustering algorithm for general use cases.

Following common clustering practices, we normalize (i.e., standardize) our parameter axes by standard deviations of the samples in their respective axes. To ensure our data is not overly noisy in the PPV space to facilitate clearer clustering, we only included samples with estimated $v_\mathrm{LSR}$ errors less than 0.085 km s$^{-1}$, i.e., the spectral resolution of our data. Given that $s_\mathrm{min}$ is the minimum neighbourhood sample required to seed a cluster, we adopt $s_\mathrm{min} = 75$ for our clustering to ensure that each cluster contains enough samples to be well resolved by the beam. We further adopt $\varepsilon = 0.17$ for our clustering, an optimal value determined from tests run with $s_\mathrm{min} = 75$. We note that the \texttt{DBSCAN} results are somewhat degenerate to the choices of $\varepsilon$ and $s_\mathrm{min}$ in that similar results can be obtained by increasing $\varepsilon$ and decreasing $s_\mathrm{min}$ slightly, and vice versa.

Our final clustering result produced under these parameters shows good agreement with structures identifiable by eye and filament-based structure identification applied to single-dish GBT NH$_3$ observations of NGC 1333 \citep{ChenMike2020}. These results are also reasonably consistent with the NGC 1333 ``ISM features'' identified by \cite{Doi2020} using \texttt{DBSCAN} on discrete N$_2$H$^+$ data. We further removed isolated pixels, also known as noise (e.g., pixels that were not grouped into clusters), and small clusters with less than 250 members from our analyses to ensure the final clusters have a minimum on-sky footprint of $\sim 6$ beams. This minimum area ensures that we have ample statistics per VCS. Since these isolated pixels and small clusters were identified independently by DBSCAN as being spatially and kinematically distinct from the larger VCSs included in this analysis, their removal will not affect our results. Our full \texttt{DBSCAN} result that includes the noise and clusters with less than 250 members is presented in Appendix \ref{appendix:DBScan}.

\begin{figure*}
    \includegraphics[width=\textwidth]{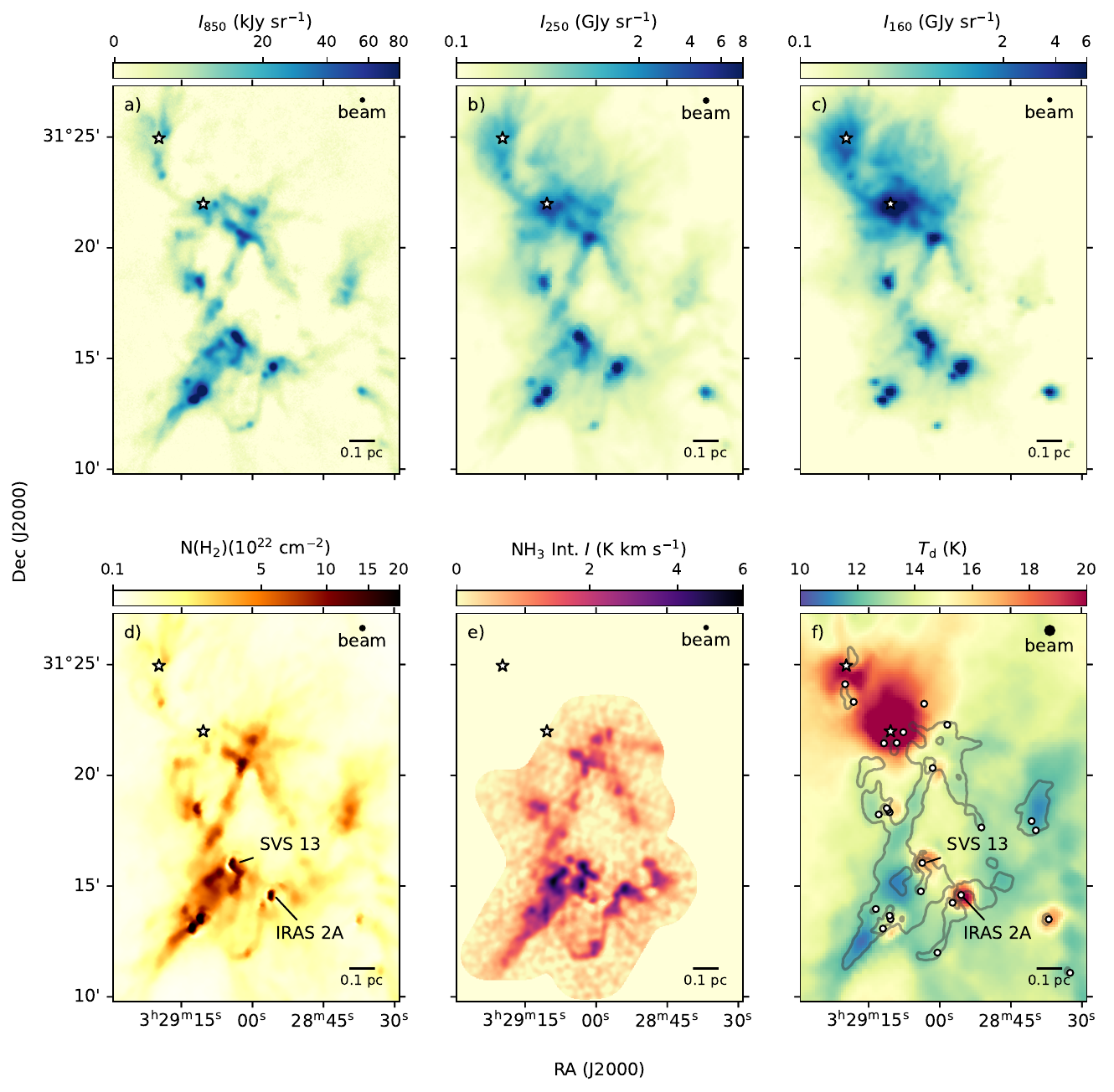}
    \caption{Observations of NGC 1333 used in this study. Top-row: a) the JGBS 850 \textmu m intensity, b) \emph{Herschel} 250 \textmu m intensity, and c) \emph{Herschel} 160 \textmu m intensity maps at their native resolutions of 14.1\arcsec, 18.2\arcsec, and 13.5\arcsec, respectively.  Bottom-row: d) the \emph{Herschel}-derived H$_2$ column density from \citet{Pezzuto2021}, e) the convolved VLA+GBT NH$_3$ (1,1) intensity integrated over a $v_{\mathrm{LSR}}$ range of $6.0 - 9.3$ km s$^{-1}$, and f) the \emph{Herschel}-derived dust temperature maps at resolutions of 18.2\arcsec, 14.1\arcsec, and 36.3\arcsec, respectively. Map resolutions for each dataset are shown in the top right corners of the plots, while 0.1 pc scale bars are shown in the bottom right corners. The locations of B-type stars \citep[e.g.,][]{Harvey1984} are indicated by the star symbols, and the positions of known Class 0/I YSOs from \citet{Dunham2015} are indicated in panel f by the open circles. The YSOs with visible column density artefacts nearby are also labelled in selective panels.}
    \label{fig:emission_maps}
\end{figure*}
%
\subsection{Gradient Calculation}
\label{subsec:grad_calc}

We denote the gradients of velocity centroid, dust emission, and column density maps as $\nabla v$, $\nabla I$, and $\nabla N(H_2)$, respectively.  Following \cite{ChenMike2020}, we calculate the local velocity, column density, and emission gradients at each pixel by fitting a planar surface using the least-square method to each map masked with a circular aperture centred on that pixel. We adopted an aperture radius of $r=14.1\arcsec$ (diameter of 28.2\arcsec) to ensure our gradients are calculated over a scale fully resolved across a diameter of two beams. While the \emph{Herschel} 250 \textmu m and $N(\mathrm{H}_2)$ maps have a slightly coarser resolution of $18.2''$, we still have at least two independent measurements with the 14.1\arcsec\ apertures.  Moreover, we tested the robustness of the \emph{Herschel} gradients using larger apertures of $r=18.2$\arcsec and found consistent results with the nominal 14.1\arcsec\ apertures. For consistency, all our gradients are calculated using the $r=14.1''$ aperture. We also only calculate gradients for apertures with more than 2/3 of their pixels with usable data.

Since the estimated errors for NH$_3$-derived $v_\mathrm{LSR}$ values can differ substantially within an aperture, our plane fitting for $v_\mathrm{LSR}$ is further inversely weighted by their estimated errors. As found by \cite{ChenMike2020}, the pixel-to-pixel variation of errors in the \texttt{MUFASA} fits, including the $v_\mathrm{LSR}$, can sometimes be significant due to a combination of factors that range from the signal-to-noise ratio (SNR) of each component to the velocity differentials between the components. Given that the emission maps' rms noise levels are fairly uniform and that the column density maps' errors are similar within an aperture, we adopted a uniform weighting to calculate the gradients of these maps.

\subsection{Quantifying Relative Orientations}
\label{sub:Z_calc}

To quantify the relative orientations (RO) between a half-vector field (e.g., the $B$ field orientation inferred from polarization) and a vector field (i.e., a gradient field), we first calculate the relative angles between these two fields at a given pixel as if they are full vectors, \textbf{a} and \textbf{b}, with the expression,
\begin{equation}\label{eq:phiPrime}
\phi' =\arccos{\frac{\mathbf{a} \cdot \mathbf{b}}{ \left|\mathbf{a}\right| \left|\mathbf{b}\right|}}.
\end{equation}
This expression yields an angle {$\phi'$} in the range of $[0, 180^{\circ}]$. A half-vector like $\mathbf{\hat{B}}$, however, is degenerate by $180^{\circ}$ in its orientation and does not distinguish between forward and backward pointings. Therefore, to account for such a degeneracy, we further map $\phi' \rightarrow \phi$ in such that all the {$\phi'$} values in the range of $[90^{\circ}, 180^{\circ}]$ are subtracted by $180^{\circ}$ and the absolute value taken. The $\phi$ resulting from this mapping is in the $[0, 90^{\circ}]$ range.

To further quantify whether a distribution of RO angles has a statistical preference for parallel alignment ($\phi=0$) or perpendicular alignment ($\phi=90^{\circ}$), we adopt the Projected Rayleigh Statistics \citep[PRS;][]{Jow2018} method for our analysis, which is a special use of the $V$ Statistics \citep[e.g.,][]{Durand1958} modified from the Rayleigh Test. Specifically, the PRS is expressed as
\begin{equation}\label{eq:PRS}
Z_{x} = \frac{\sum_{i}^{n} \cos{(2\phi_i)}}{\sqrt{n/2}},
\end{equation}
for a set of $\phi_i$ angles with a size $n$. A $\phi$ distribution with $Z_{x} > 0$ or $Z_{x} < 0$ shows a statistical preference for a parallel or perpendicular alignment, respectively. To quantify the significance of the $Z_{x}$ values, we estimate the $Z_{x}$ variance as
\begin{equation}\label{eq:PRS_variance}
\sigma^2_{Z_{x}} = \frac{2\sum_{i}^{n} [\cos{(2\phi_i)}]^2 - [Z_x]^2}{n},
\end{equation}
and adopt $\sigma_{Z_{x}}$ as our estimated uncertainty that reflects the dispersion in $\phi$ \citep{Jow2018}. To ensure $Z_{x}$ is calculated from reasonably independent measurements, we only take one $\phi_i$ sample per beam in our PRS calculations.

Since we are analyzing the relative angles between the $B$ field and several different gradient fields, we will omit the $x$ subscript from the $Z_{x}$ whenever additional subscripts are used to specify to which relative angles the $Z_{x}$ is referring. Specifically, $Z_{x}$ calculated from the relative angles between the $B$ field and the $\nabla\,v$, $\nabla N(\mathrm{H}_2)$, and $\nabla I_\lambda$ fields will be designated $Z_{Bv}$, $Z_{BN}$, and $Z_{BI}$, respectively. We note that \cite{Jow2018} computes their $Z_{x}$ between column density gradients and dust polarization, rather than magnetic fields, to be consistent with the analyses of \cite{Planck2016_HROMagClouds} and \cite{Soler2017}. In these works, the convention is to measure the ROs between the magnetic fields and ``structure alignments'' (i.e., isocolumndensity contours that are perpendicular to column density gradients). The $Z_{x}$ values computed by \cite{Jow2018} will thus have an opposite sign than our $Z_{x}$ values.

\section{Results}
\label{sec:results}


\subsection{Emission and Column Density Maps}
\label{subsec:emission_maps_results}

Figure \ref{fig:emission_maps} shows the data of NGC 1333 used in this analysis. The top panels give the 850 \textmu m, 250 \textmu m, and 160 \textmu m dust emission maps, and the bottom panels show the $N(\mathrm{H}_2)$, NH$_3$ (1,1) integrated intensity, and $T_d$ maps. Overall, the 850 \textmu m and the $N(\mathrm{H}_2)$ maps appear morphologically similar. As the dust emission wavelength decreases, however, extended emission becomes more significant. In particular, we see pronounced diffuse structures in the 250 \textmu m and 160 \textmu m map in the northeast region near the two B stars. This extended dust emission also appears to be warm based on the $T_d$ map, indicating that the B stars are heating the surrounding material. The $T_d$ map also shows compact warm structures at the positions of the Class 0/I YSOs (see Figure \ref{fig:emission_maps}f) identified by \cite{Dunham2015}.

Aside from the missing larger-scale emission, the 850 \textmu m and NH$_3$ maps resemble the $N(\mathrm{H}_2)$ map reasonably well, particularly compared to the 250 \textmu m and the 160 \textmu m maps. The 850 \textmu m data is insensitive to large-scale emission due to spatial filtering during the data reduction process, required for atmospheric emission removal. While the NH$_3$ map was imaged with feathered data between the VLA and GBT observations to ensure it is sensitive to all scales, NH$_3$ is not typically detected in diffuse low-density ($< 10^3$ cm$^{-3}$) gas because (1) this gas tends to be warmer resulting in more carbon-rich gas chemistry that disfavours NH$_3$, and (2) lower density gas typically does not have the excitation conditions for NH$_3$ \citep[e.g.,][]{BerginTafalla2007, Shirley2015}.

\begin{figure*}
    \includegraphics[width=0.86\textwidth]{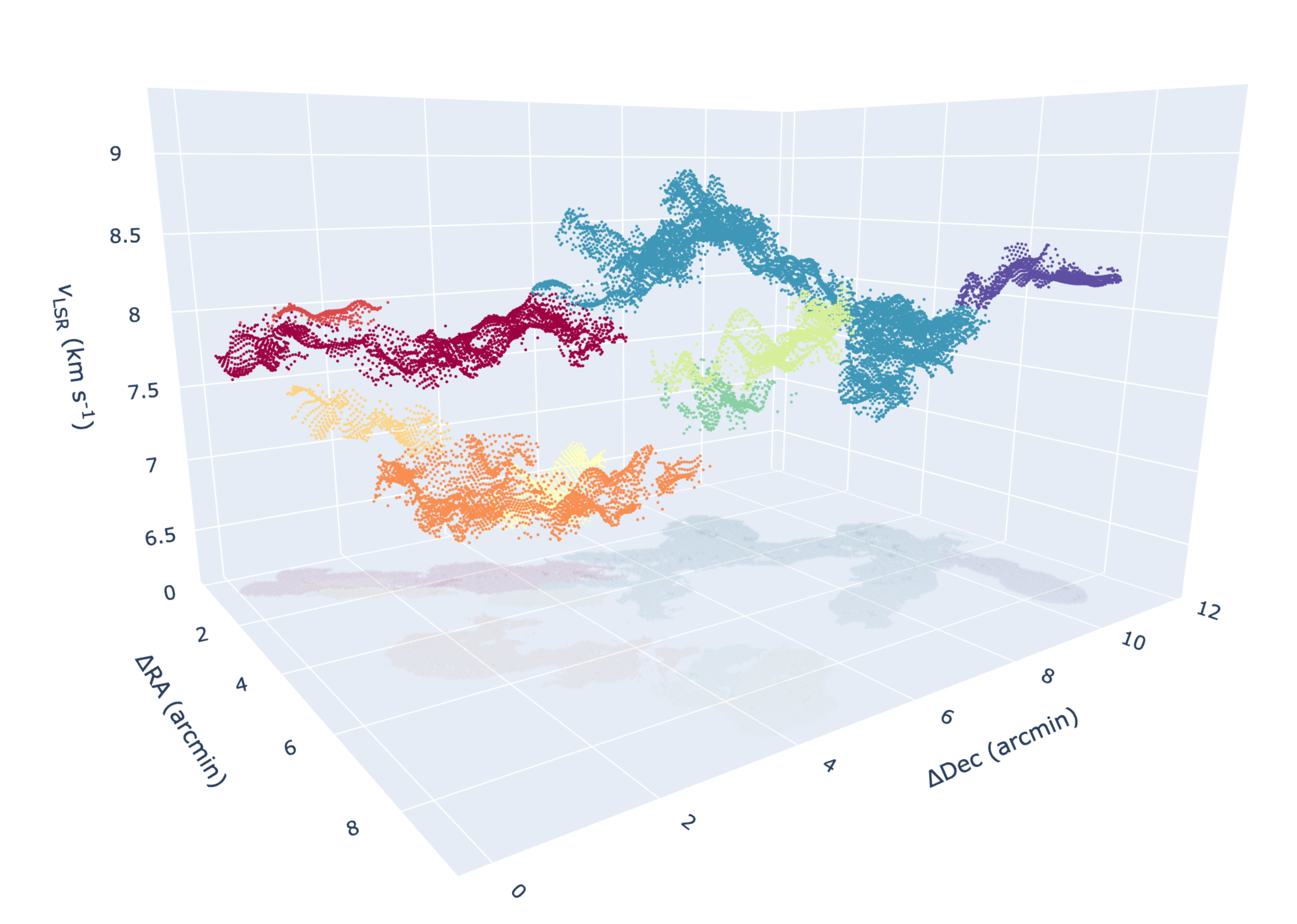}
    \caption{A 3D scatter plot of the nine identified NGC 1333 VCSs we identified from the fitted NH$_3$ spectral models using \texttt{DBSCAN} in PPV space. Each VCS is shown in a different colour.  The data points show the corresponding PPV pixels for identified members, using the $v_\textrm{LSR}$ for the velocity. The plane-of-the-sky projection of the VCSs is represented as `shadows' on the RA-decl. plane using the same colours. The position coordinates are relative to the lower-left corner of the map.}
    \label{fig:dbscan_3d}
\end{figure*}

\subsection{Velocity-Coherent Structures}
\label{sub:vcs_result}

Following Section \ref{sec:methods}, we fitted the NH$_3$ spectra to probe the gas kinematics in NGC 1333 and identify VCSs. We detected NH$_3$ spectra in 45\% of the pixels over the area mapped by the GBT and VLA. Out of these detections, 33\% are better fitted with two-component models, indicating that multiple-component fits are important for NH$_3$ in NGC 1333, consistent with the single-dish findings of \cite{ChenCY2020}.

Figure \ref{fig:dbscan_3d} shows a 3D scatter plot of the nine VCSs we identified in NGC 1333 in PPV space using \texttt{DBSCAN} (see Section \ref{subsec:vcs_def}). Each VCS is shown by a different colour, and they appear to be fairly kinematically distinct from one another. Similar to studies that looked at gas kinematics of star-forming filaments in PPV space or PV (position-velocity) projections, both in observations (e.g., \citealt{Hacar2013}) and simulations (e.g., \citealt{Clarke2016}), we see quasi-oscillatory behaviour in the $v_\mathrm{LSR}$ within these VCSs. Such a behaviour is also consistent with that found by \cite{ChenMike2020} in NGC 1333 with single-dish GBT NH$_3$ data. 

\begin{figure*}
    \includegraphics[width=\textwidth]{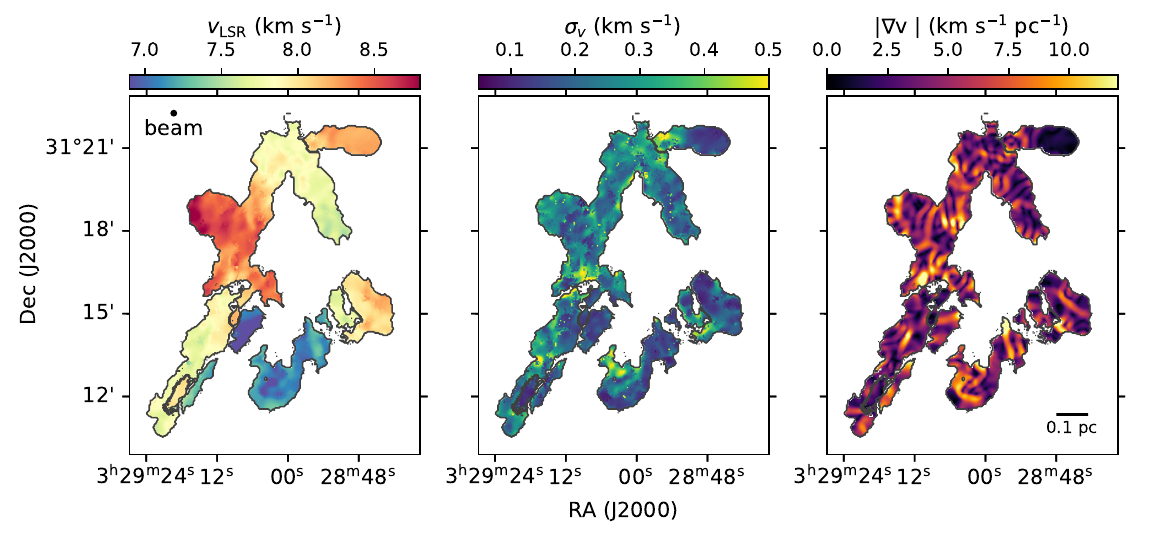}
    \caption{Maps of $v_{\mathrm{LSR}}$ (left), $\sigma_{v}$ (middle), and $|\nabla v|$ (right) for the nine identified VCSs in NGC 1333 using the NH$_3$ (1,1) data at 14.1\arcsec\ resolution. The VCSs are separated by black contours to represent their boundaries. We note these VCSs have minimal spatial overlaps on the plane of sky due to most of the faint, second velocity components being excluded from our analyses because of their large $v_{\mathrm{LSR}}$ errors (see text for details). The small black shapes near the edges of the VCSs are examples of the isolated pixels and small clusters that were excluded from this analysis.
    }
    \label{fig:kinMaps}
\end{figure*}

Figure \ref{fig:kinMaps} shows maps of $v_\mathrm{LSR}$, $\sigma_v$, and velocity gradient ($|\nabla v|$) for NGC 1333 in the left, centre, and right panels, respectively.  We evaluate these quantities separately for each of the nine VCSs. Since we only included fitted velocity components with estimated $v_\mathrm{LSR}$ errors greater than 0.085 km s$^{-1}$ (see Section \ref{subsec:vcs_def}), most of the second components from our fits are not present in our clusters due to their larger $v_\mathrm{LSR}$ errors. The overlap of these VCSs on the plane of the sky is thus minimal. We note that while the fitting error for a faint, second velocity component may be large, two-component fits are still needed to capture the observed spectra adequately and derive gas properties accurately, even for the bright component \citep[e.g.,][]{ChenMike2020, Choudhury2020}.

Visually, $v_\mathrm{LSR}$ is smooth across each VCS, indicating these VCSs are indeed velocity-coherent. The $v_\mathrm{LSR}$ differences between VCSs and their immediate neighbours can be quite large, consistent with what we see in the PPV space (see Figure \ref{fig:dbscan_3d}). The $|\nabla v|$ values of each VCS are not elevated near the edges, further demonstrating that the gradients we calculated are purely internal to the VCSs themselves (i.e., intra-VCS) and are free from contamination from velocity components belonging to other VCSs.

The $|\nabla v|$ maps of the VCSs in NGC 1333 contain quasi-periodic ripples. Such structures were seen previously in the NH$_3$ observations of Perseus B5 at higher resolutions ($5''$; \citealt{ChenMike2022}) and in NGC 1333 using lower-resolution ($32''$) single-dish data \citep{ChenMike2020}. Similar wave-like structures have also been implied in the  $v_\mathrm{LSR}$ profiles along filament spines observed with other tracers (e.g., CO; \citealt{Hacar2013}) and with synthetic observations of simulations (e.g., \citealt{SmithR2016}). These earlier studies, however, did not explore these oscillations in 2D beyond filament spines. 

The $|\nabla v|$ ripples may be gravo-acoustic in nature, like those found in simulations by \cite{Clarke2016}, or driven by MHD waves, such as those proposed by \cite{Tritsis2016}. In the case of magnetosonic waves, the waves are expected to travel perpendicularly to the magnetic field. We will discuss these $|\nabla v|$ ripples further in Section \ref{sec:discussion}, particularly in \ref{subssec:Z_vgrad}. 

\subsection{The Magnetic Field and Velocity Gradients}
\label{sub:BnVGrad_result}

\begin{figure*}
    \includegraphics[width=0.95\textwidth]{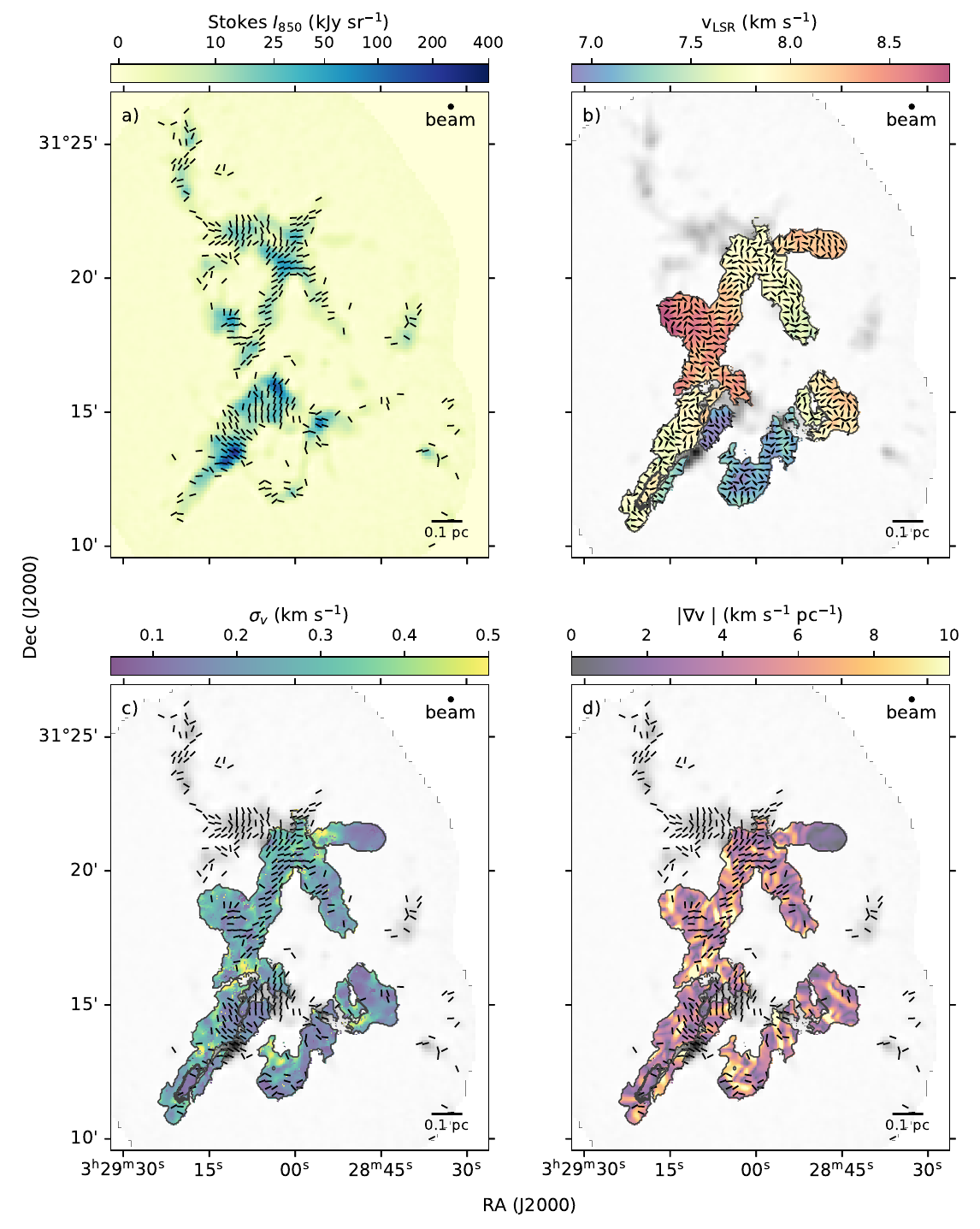}
    \caption{The inferred magnetic field orientation from the BISTRO polarization data of NGC 1333 shown as line segments (polarized half-vectors rotated 90$^\circ$), overlaid on the corresponding BISTRO Stokes $I$ map and the $\sigma_v$ and $|\nabla v|$ maps of the VCSs in panels a, c, and d, respectively. Only one segment per 14.1\arcsec\ beam is shown here. The normalized $\nabla v$ fields of each VCS are shown in panel b as stylized arrows, with one arrow per beam, overlaid on the $v_{\mathrm{LSR}}$ maps from which they are calculated. The BISTRO 850 \textmu m Stokes I data shown in panel a are further displayed in the backgrounds of panels b, c, and d in greyscale.}
    \label{fig:field_maps}
\end{figure*}

Figure \ref{fig:field_maps} shows the inferred magnetic field as half-vectors over the 850 \textmu m continuum, $\sigma_v$ and $|\nabla v|$ maps for the VCSs in panels a, c, and d, respectively. Panel b shows the velocity gradient field as full vectors on top of each VCS's $v_\mathrm{LSR}$ maps, with the 850 \textmu m continuum map further placed in the background. These data showcase a lack of NH$_3$ detection toward the northeast corner of NGC 1333, typically above decl. $\sim 31^{\circ} 22'$. While NH$_3$ observation in that region is predominately lacking due to the limited footprint of the VLA mosaic, we still have partial coverage near the southern B star that show none-detection of NH$_3$ that is consistent with the GBT-only observations \citep[e.g.,][]{GAS}. The lack of NH$_3$ emission towards there is likely due to a change in the gas chemistry or excitation conditions of the gas itself. Specifically, this region of NGC 1333 is likely being heated from the nearby B stars \citep[e.g., see Fig. \ref{fig:emission_maps}f and][]{ChenMike2016}. Such heating can release carbon-bearing molecules back into the gas phase and deplete NH$_3$ chemically \citep{BerginTafalla2007}. Furthermore, since the H$_2$ column densities toward the B stars are lower, even if we assume a similar NH$_{3}$ abundance, the H$_2$ spatial densities behind these column densities may be lower than NH$_{3}$'s critical excitation value. Indeed, \citet{Doi2020} found a similar lack of N$_2$H$^+$ emission in this region, suggesting the mechanisms behind the null-detection are similar. We note that higher dust temperatures can cause the 850 \textmu m dust emission to be brighter in that region than expected from its column density (see Fig. \ref{fig:emission_maps}), making polarized emission easier to detect for a given polarization fraction. We note that the NH$_3$ observations do not cover areas west of R.A. $\sim3^\mathrm{h}28^\mathrm{m}40^\mathrm{s}$. 

As seen in Figure \ref{fig:field_maps}d, the dust polarization in NGC 133 tends to be better detected towards lower $\sigma_v$ regions, where the column density also tends to be higher. The trend where density anticorrelates with $\sigma_v$ is consistent with those seen in NH$_3$ observations, either as a sharp sonic transition (e.g., \citealt{Pineda2010}) or gradual decrease within subsonic structures (e.g., \citealt{ChenMike2022}). Broadly, higher-density structures have lower turbulence themselves, likely due to them being harder to perturb than their lower-density counterparts \citep[e.g.,][]{Heigl2020}.

Figure \ref{fig:field_maps}d demonstrates that the $|\nabla v|$ ripples noted previously appear to be aligned perpendicular to the $B$ field in some places and parallel in others. Considering that slow and fast MHD waves (i.e., Alfv\'{e}n and magnetosonic waves) propagate in directions that are along and perpendicular to the unperturbed magnetic field lines, respectively \citep{Tritsis2016}, the $|\nabla v|$ ripple may indeed trace the motions of both MHD waves in our VCSs. We will discuss these ripples further in Section \ref{subssec:Z_vgrad}.

\subsection{Relative Orientation Angles}
\label{sub:RO_result}

\begin{figure*}
    \includegraphics[width=\textwidth]{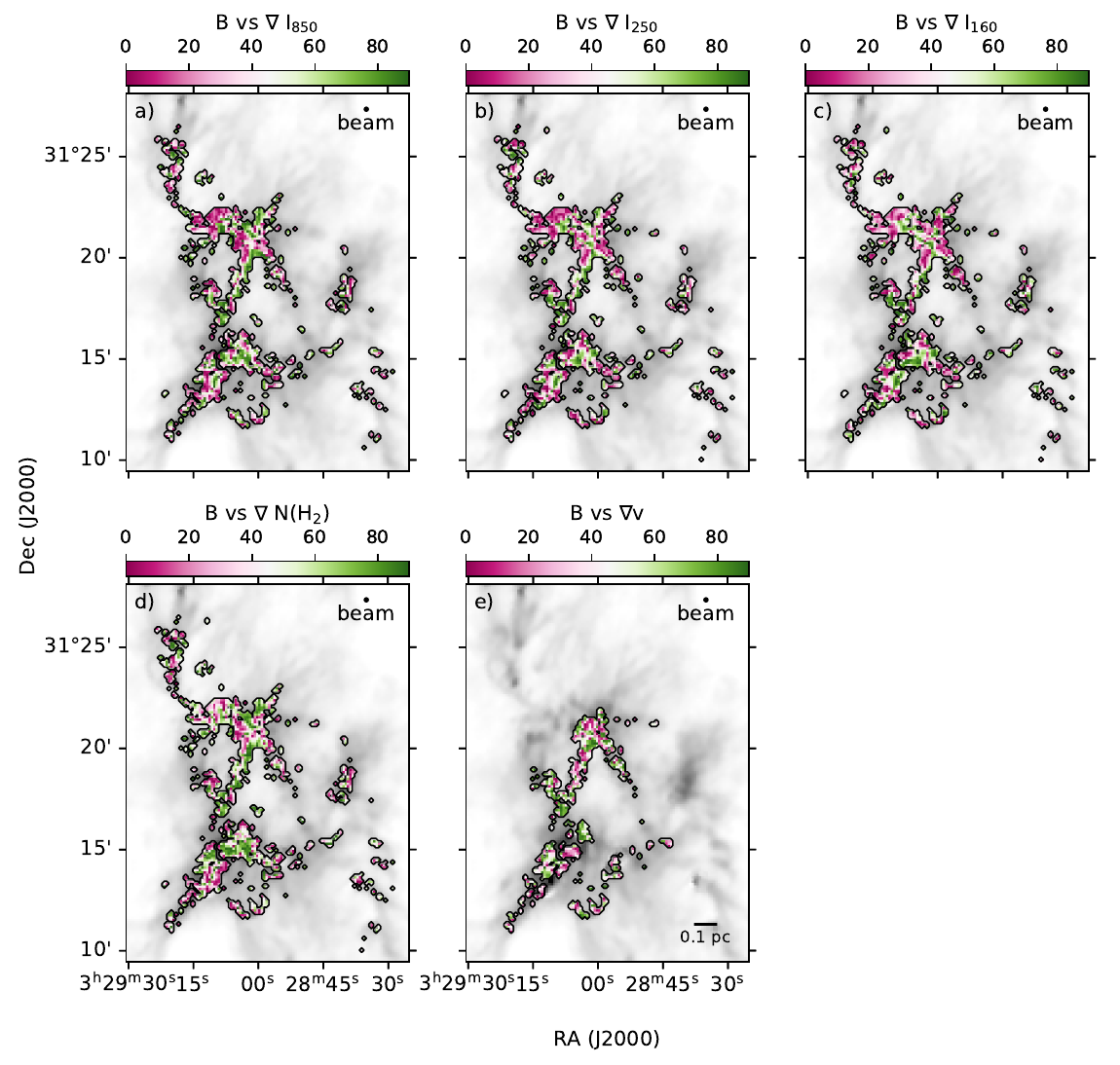}
    \caption{Maps of RO across NGC 1333 as measured between the inferred $B$ field and the gradient fields, i.e., a) $\nabla I_{850}$ b) $\nabla I_{250}$, c) $\nabla I_{160}$, d)$\nabla N(\mathrm{H}_2)$, and e) $\nabla v$ over the same \emph{Herschel}-derived $N(\mathrm{H}_2)$ map from Figure \ref{fig:emission_maps} in greyscale. The black contour shows the region for which measurements of the RO were made. ROs of 0$^\circ$ and 90$^\circ$ indicate parallel and perpendicular alignments between these fields, respectively (see text for further details).
    \label{fig:ROMaps}}
\end{figure*}

In this section, we present the RO angles (i.e., $\phi$) between the $B$ field and the different gradients in NGC 1333. Specifically, we compare the $B$ field's orientation to the velocity, emission, and column density gradients using a convention where $0^{\circ}$ and $90^{\circ}$ reference parallel and particular alignments, respectively. We note that our convention differs from similar RO studies in the literature by 90$^\circ$, where the angle is defined between the polarization position angle and the column density gradient that equivalently measures alignments between the $B$ field and the cloud structure elongations \citep[e.g.,][]{Planck2016_HROMagClouds,Jow2018}. In other words, parallel gradient alignment in our convention is perpendicular to the other conventions and vice versa. Nevertheless, we use this convention to apply the same analysis to the emission and column density gradients as the velocity gradients.

\subsubsection{Global Distributions}
\label{subsub:global_trends}

Figure \ref{fig:ROMaps} shows maps of RO angles between the $B$ field and the gradients of dust emissions (for the 850 $\mu$m, 250 $\mu$m, and 160 $\mu$m data), H$_2$ column density, and NH$_3$ velocity centroid. Hereafter, we refer to the gradients in either the thermal dust emission or the column density maps as structure gradients and the gradients in the $v_\mathrm{LSR}$ as velocity gradients.   

As can be seen, the ROs are neither globally uniform nor slowly varying across the entire clump. These ROs, however, are not completely random either and appear locally coherent, with many regions showing preferential parallel or perpendicular alignments. The ROs of the structure gradients (i.e., the emission and column density gradients) share similar localized features with each other but not with their velocity gradient counterpart. 

\begin{figure*}
    \includegraphics[width=0.8\textwidth]{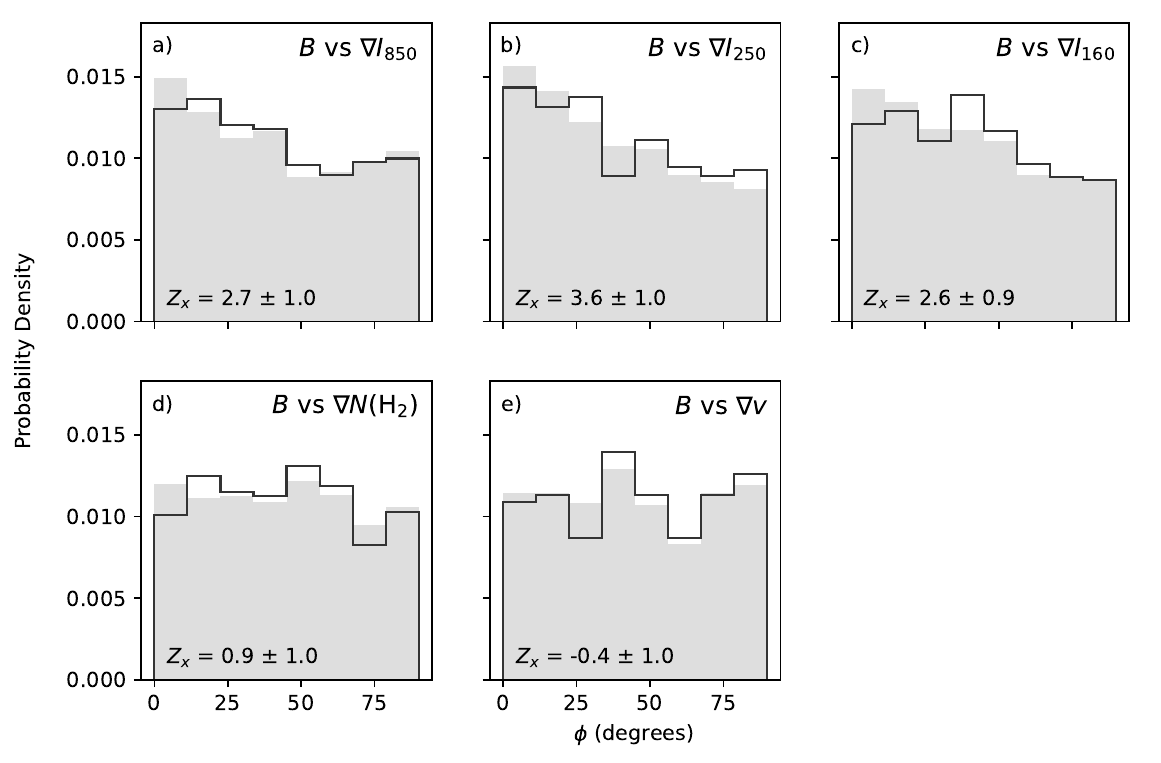}
    \caption{Histograms showing the RO ($\phi$) distributions in NGC 1333 between the magnetic field and the four structure gradients (i.e., $\nabla I_{850}$, $\nabla I_{250}$, $\nabla I_{160}$, $\nabla N(\mathrm{H}_2)$) and the velocity centroid gradient ($\nabla v$).  All histograms have been normalized to show probability density functions.  The grey-filled histograms show the distribution of all the pixels, while the black-open histograms show the results for independent measurements only (e.g., one measurement per beam). The Projected Rayleigh Statistics (PRS) $Z_x$ and for the independent measurements (open histograms) and its 1 $\sigma$ uncertainty (see Section \ref{sub:Z_calc}) are given in the bottom-left corners of each panel.}
    \label{fig:global_histo}
\end{figure*}

Figure \ref{fig:global_histo} shows histograms of RO distributions between the $B$ field and each of the structure and velocity gradients. The grey-filled and black open histograms show the RO distribution for all measurements (found within the black contours in Figure \ref{fig:ROMaps}) and those from independent, beam-separated measurements, respectively. As described in Section \ref{sub:Z_calc}, only independent measurements are used to calculate the PRS $Z_\mathrm{x}$ values. The  $Z_\mathrm{x}$ values for each histogram are given in the bottom-left corners of each panel.  Both $B$ vs. $\nabla v$ and $B$ vs. $\nabla N(\mathrm{H}_2)$ have $Z_\mathrm{x}$ values consistent with zero within the measured $1\sigma$ uncertainties, indicating there is no global preference for parallel or perpendicular alignments.  

By contrast, the RO distributions between $B$ and the dust emission gradients (i.e.,  $\nabla I_{850}$, $\nabla I_{250}$, $\nabla I_{160}$) appear preferentially parallel.  In all three cases, $Z_x$ is positive and inconsistent with zero above $> 2.5\sigma$.  This result agrees well with larger-scale ($10'$; $\sim 0.9$ pc for Perseus) observations where the magnetic field aligns parallel with structure gradients at high densities that encompass our entire data samples \citep[e.g.,][]{Planck2016_HROMagClouds}. Since emission gradients are generally orthogonal to the direction of elongated structures, these results translate to a perpendicular \textit{structure alignment}. On cloud scales, such an alignment has been interpreted as gas contraction and accretion along the magnetic field \cite{Soler2017}. A similar process may, therefore, explain filament formation at the scale we probe, which are two orders of magnitudes smaller.  We will discuss the implication of these global alignments in Section \ref{subsec:global_trends_discuss}.  Nevertheless, we see pockets of both parallel and perpendicular ROs across NGC 1333 (see Figure \ref{fig:ROMaps}) that indicate the localized cloud properties are also important. In the next subsections, we examine the ROs for subregions of NGC 1333.

\subsubsection{Local Velocity Gradients}
\label{subsec:local_vgrad}

\begin{figure*}
    \includegraphics[width=0.84\textwidth]{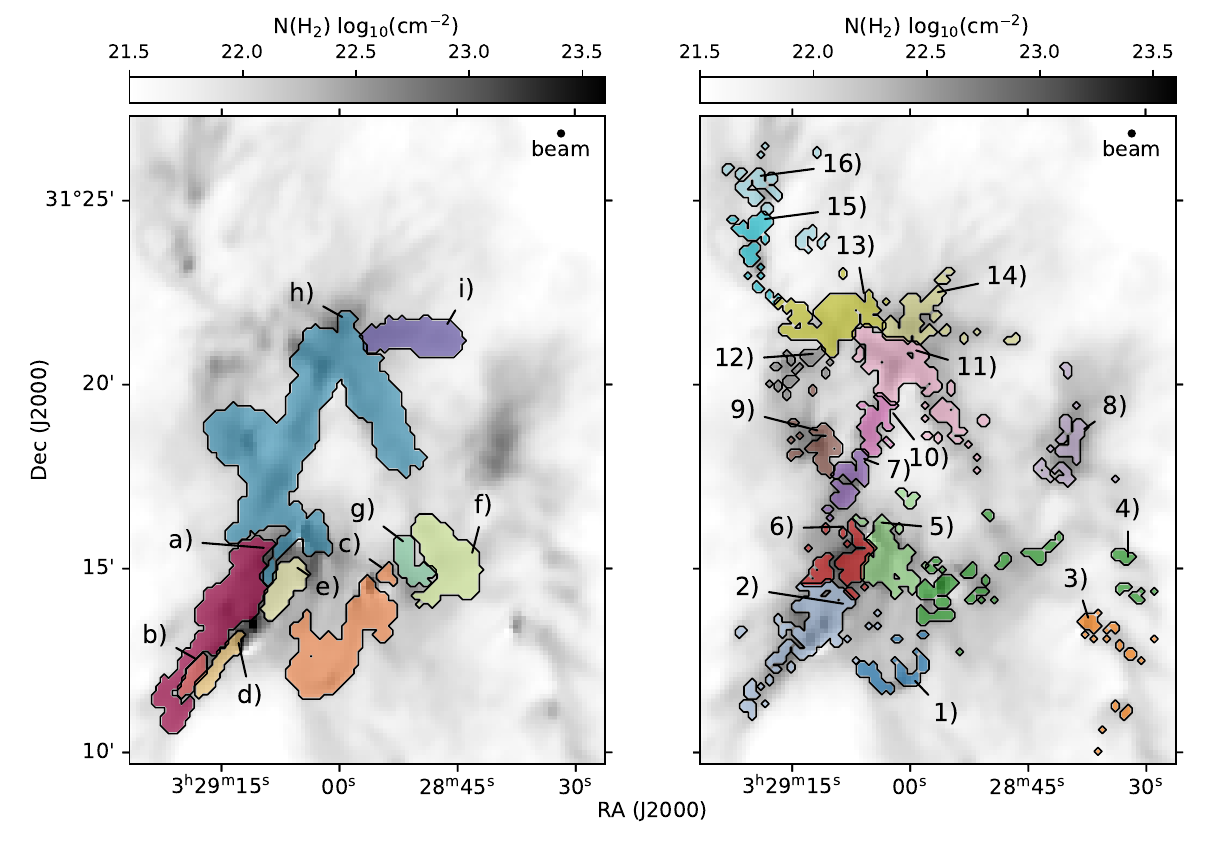}
    \caption{Definition of subregions in NGC 1333 in colour with the same \emph{Herschel}-derived $N(\mathrm{H}_2)$ map from Figure \ref{fig:emission_maps} in greyscale. Left: The VCSs identified from \texttt{DBSCAN} clustering. Right: Subregions determined by using the watershed method on the 850 \textmu m data, for only where $B$ field is robustly detected. Each subregion is indicated by a unique colour and label. }
    \label{fig:regions}
\end{figure*}

\begin{figure}
    \includegraphics[width=\columnwidth]{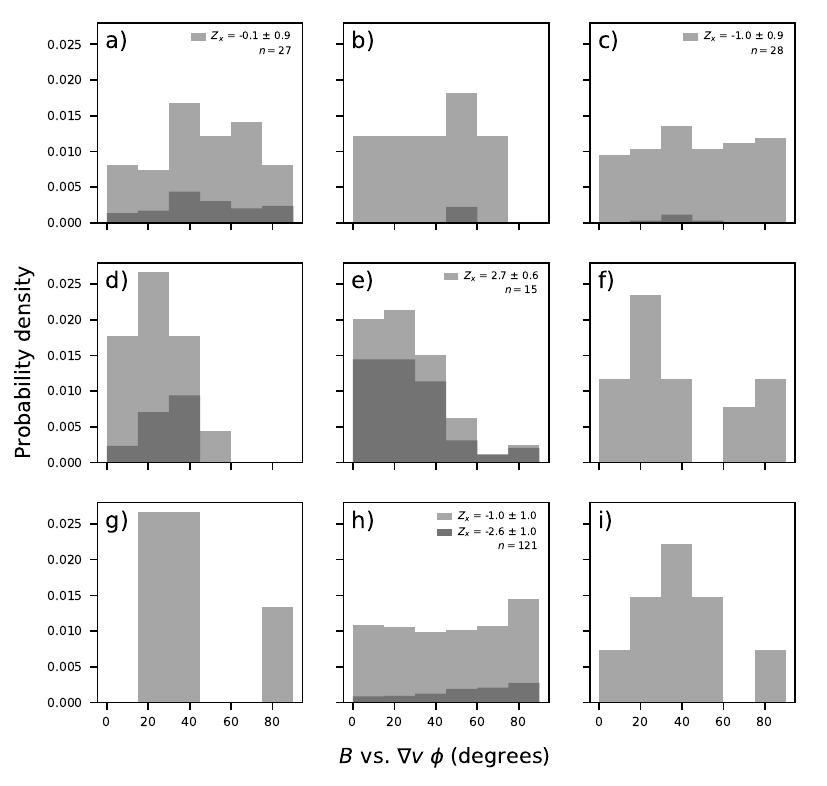}
    \caption{Normalized histograms of RO between $B$ field and $\nabla v$ for each of the VCSs (see left panel of Fig. \ref{fig:regions} for labels). The light and dark grey histograms represent all the RO measurements and those taken at $N$(H$_2$) $\geq 5\times 10^{22}$ cm$^{-2}$, respectively. $Z_\mathrm{x}$ values of distributions with $\geq 14$ independent measurements are shown in the top-right corner, accompanied by the number of independent measurements for the entire region.}
    \label{fig:HRO_vcs}
\end{figure}

Figure \ref{fig:regions} (left) shows the nine VCSs identified in NGC 1333 using \texttt{DBSCAN} (see Section \ref{subsec:vcs_def}), labelled with letters.  We use these labels to identify each VCS in the subsequent analysis.   Figure \ref{fig:HRO_vcs} shows the RO distributions for each of the VCSs, where the light and dark grey histograms represent all pixels and pixels with $N$(H$_2$) $\geq 5\times 10^{22}$ cm$^{-2}$, respectively. Most VCSs do not sufficiently overlap with robust $B$ measurement for reliable PRS calculations. Therefore, we only report the $Z_\mathrm{x}$ values for four (out of nine) VCSs with $\geq 15$ independent measurements in Figure \ref{fig:HRO_vcs}.

Visually, Figure \ref{fig:HRO_vcs} shows that most VCSs have no preferred alignments between $B$ and $\nabla v$.
\textit{VCS-e} appears to be an exception, with a positive $Z_\mathrm{x} > 2.5\sigma$ that indicates the magnetic field is parallel to the velocity gradient in this region. This result is consistent with the qualitative alignment found by \citetalias{Dhabal2019} over the same area of NGC 1333 using NH$_3$-derived velocity gradients with the $B$-field presented by \cite{Doi2020}. 
\citetalias{Dhabal2019} argued that a parallel $B$--$\nabla v$ alignment, in addition to the $B$-field running perpendicular to the dense structure, indicates that this region is likely formed by large-scale compression due to an expanding bubble. \citetalias{Dhabal2019}, however, only performed a single-velocity fit to their $4''$ NH$_3$ data and treated all of NGC 1333 as a single coherent region. As a result, their velocity gradients correspond to a sharp jump between the nearby subregions we identified, \textit{VCS-a} and \textit{VCS-d}, and therefore may not be tracing a kinematically continuous structure. We investigate further the implications of mass flow in compressed, magnetized gas at this region in Section \ref{subsec:postshock_flow}.

\subsubsection{Local Structure Gradients}
\label{subsub:local_sgrad}

We divide NGC 1333 into emission-based regions instead of kinematic-based VCSs for structure gradient analyses. This approach allows us to examine better the local orientations between the $B$ field and the various structure gradients. We employ the watershed method \citep[e.g.,][]{Beucher1993} to delineate our regions, which draws boundaries between seeded peaks at locations where the emission structures first meet, marking topological features akin to valleys. We specifically ran the watershed method on the JGBS 850 \textmu m emission map to define structures observed in the same wavelength and with the same facility as the $B$ field data. The right panel of Figure \ref{fig:regions} shows the 16 watershed regions identified in NGC 1333 over pixels where the $B$ field is robustly detected. 

\begin{figure}
    \includegraphics[width=\columnwidth]{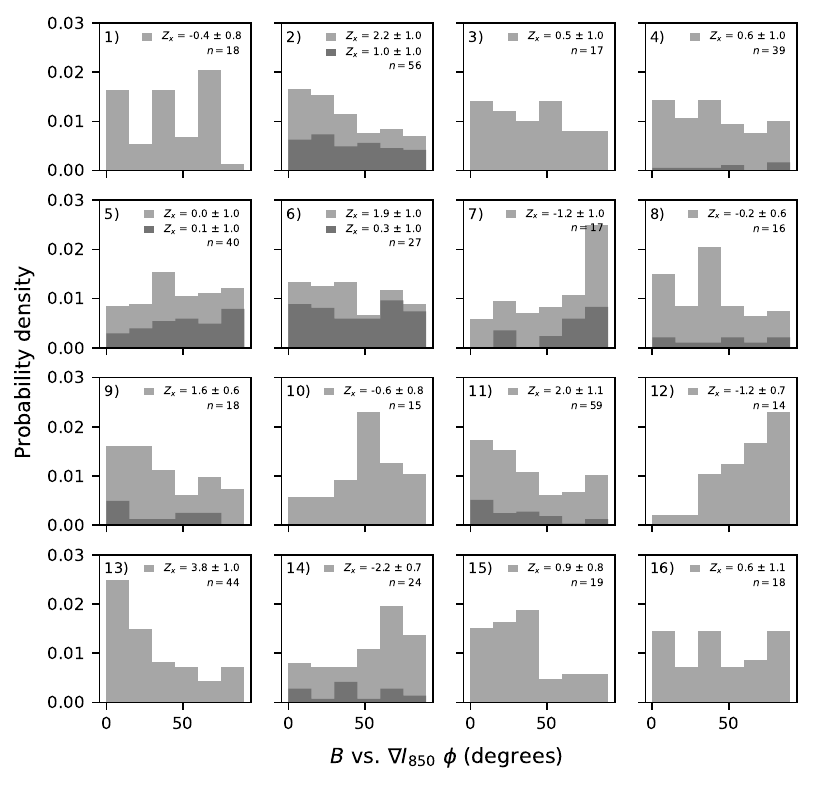}
    \caption{Normalized histograms of RO distributions for the $B$ field and $\nabla I_{850}$ for each of the watershed regions in NGC 1333 (see right panel of Fig. \ref{fig:regions} for the labels). The dark and light grey histograms, as well as the conventions for the annotated $Z$ values, are the same as in Figure \ref{fig:HRO_vcs}. }
    \label{fig:HRO_watershed}
\end{figure}

Figure \ref{fig:HRO_watershed} shows the RO distributions between $B$ and $\nabla I_{850}$ for each of the 16 watershed regions, with their corresponding PRS $Z_\mathrm{x}$ values and errors. The PRS values are calculated from independent measurements only and their corresponding sample size ($n$) is also given. When examined by eye, some distributions appear to have preferential alignment. For example, \textit{Region-2} and \textit{Region-13} appear to have preferentially parallel alignment, whereas \textit{Region-7} and \textit{Region-12} appear to have preferentially perpendicular alignment. Most of the subregions, however, do not have $|Z_x|$ significantly above zero, partially due to the limited number of independent RO measurements in most regions. The only regions with robust ($Z_{x} >2.5 \sigma$) PRS measurements are \textit{Region-9} and \textit{Region-13}, which have preferential parallel alignment, and \textit{Region-14}, which shows perpendicular alignment. \textit{Region-2} has a tentative ($> 2\sigma$) preference for parallel alignment that could become more robust with increased sample size. We note that PRS is insensitive to preferential alignment for $\phi \sim 45^{\circ}$. Indeed, \textit{Region-10} shows preferential $\sim 45^{\circ}$ alignment, similar to that found by \cite{Doi2020} in the same region relative to the filament's position angle, and has $|Z_x| < 1 \sigma$. Nevertheless, distribution like that of \textit{Region-10} seems rare in NGC 1333. Since the RO distributions between $B$ and the other emission gradients in these watershed regions behave similarly to the 850 \textmu m results, we will not present those results here. 

\textit{Region-13}, \textit{15}, and \textit{16} are located in the vicinity of the nearby B-stars and may, therefore, be influenced by them (see Fig. \ref{fig:emission_maps}). Stellar feedback from the B-stars can heat dust substantially and potentially alter magnetic field morphology. Interestingly, \textit{Region-13} shows preferential parallel alignment, while \textit{Region-15} and \textit{16} have $Z_x$ values consistent with zero, indicating no preferential alignment. We note, however, that the PRS for \textit{Region-15} and \textit{16} may be limited by small sample sizes. Even though \textit{Region-13} has the highest median dust temperature in NGC 1333 ($\sim 20$ K), \textit{Region-15} and \textit{16} are somewhat cooler in comparison ($\sim  18$ K and $\sim  14$ K, respectively) and may be better shielded from the heating. Indeed, \textit{Region-15} and \textit{16} both contain cool pockets of local overdensities in an otherwise warm, B-star-heated region (see Fig. \ref{fig:emission_maps}). Similarly, NH$_3$ is detected towards denser parts of \textit{Region-15} and \textit{16} but not in \textit{Region-13} based on GAS observations \citep{GAS} which cover the north-east quadrant of NGC 1333 unlike the data presented here. The non-detection of NH$_3$ in \textit{Region-13} from the GAS data is consistent with that expected for a warm, low column density region (see also, panels d and f of Figure 1) due to the gas chemistry and excitation conditions there.

We examined the relationship between $Z_\mathrm{x}$ and the median $N(\mathrm{H}_2)$, $T_\mathrm{d}$, $|\nabla v|$, and $\sigma_v$ for each subregion, but found no clear correlation. By contrast, we find correlations when binning the RO measurements by gas properties instead of defined subregions (see Section \ref{subsec:gas_properties}).  This distinction suggests that our watershed regions contain a mix of gas populations that erases any potential alignment trends on these scales when averaged. For example, many of these regions can contain both pre- and protostellar sources, which can have significantly different dust temperatures.  We discuss the effect of gas properties further in Section \ref{subsec:gas_properties}.

\section{Discussion}
\label{sec:discussion}

\subsection{Global Alignments and Structure Formation}
\label{subsec:global_trends_discuss}

In Section \ref{subsub:global_trends}, we found that the magnetic field and the emission gradients are preferentially aligned in parallel on average in NGC 1333. In terms of the convention more commonly expressed in the literature, i.e., structure alignment, our results indicate that the magnetic field is preferentially aligned perpendicular to the density structures themselves, similar to those seen on larger scales by \cite{Planck2016_HROMagClouds}. Such an alignment has been attributed to structures forming via gas contraction and accretion along the magnetic field \citep{Soler2017}. Similarly, MHD simulations by \cite{ChenCY2015}, which look at filament and magnetic field alignments on $\sim 0.05$ pc scales comparable to our study, also found preferential perpendicular alignment at a gas density $\gtrsim 2 \times 10^{22}$ cm$^{-2}$ \citep{ChenCY2016}, consistent with our results. In these simulations, dense structures (e.g., filaments and cores) were initially seeded as overdensities in post-shock regions and subsequently grew through gas accretion along the magnetic field lines. The transition from parallel to perpendicular structure alignment occurs at a regime where the kinetic energy of the gas structure dominates over the magnetic support \citep{ChenCY2016}. The resulting gravitational contraction in this regime then distorts the magnetic field, causing more perpendicular alignment between the POS magnetic field and the gas structures.

\cite{Doi2020} found no global structure alignment between the magnetic field and filament position angles in NGC 1333 using the same polarization data.  Nevertheless, they did find these relative alignments to be fairly uniform within individual filaments and demonstrated that the magnetic field orientation could be perpendicular to the filaments in 3D relative and only appear misaligned due to varying inclinations. Using geometric models, \citeauthor{Doi2020} showed that the magnetic field should be preferentially perpendicular to the density structure, even for filaments randomly oriented in 3D. Our results here are thus consistent with such a simple geometric model, where filaments are assumed to align perpendicularly to the magnetic field in 3D but have random inclination angles relative to the lines of sight.

The global RO distribution between $B$ and $\nabla N$(H$_2$) does not show the same trends as their emission gradient counterparts. This discrepancy is initially unexpected, given that the 160 \textmu m, 250 \textmu m, and 850 \textmu m dust emissions are expected to approximate the column densities well to the first order, and suggests that higher-order effects are influencing the column density map relative to the emission maps. For example, dust emission and polarization maps can be sensitive to warmer structures, particularly at shorter wavelengths. Optically thin dust emissions and their polarized components can thus be weighted more by temperature than column density along the lines of sight. This higher-order effect likely explains why the trends between the $B$ field and the emission gradients are not found in their column density counterpart. We will discuss the evidence for temperature bias further in Section \ref{subsub:Z_dust_temp}, where we find emission structures to align more perpendicularly with the magnetic field at higher dust temperatures. Appendix \ref{appsub:ColDen_v_Emission} also demonstrates how the column density map resembles the dust emission maps, particularly 850 \textmu m, at low temperatures and less so as dust temperature increases.

In addition to temperature biases, we note that the $18.2''$ column density map adopted in this paper was derived assuming a fixed dust opacity spectral index ($\beta$) using the $36''$ \textit{Herschel}-derived dust temperature map \citep{Pezzuto2021}. The $\beta$ values in NGC 1333 near protostars and outflows, however, tend to be lower than those assumed for the \textit{Herschel} column density maps \citep{ChenCY2016}. The $36''$ $T_\mathrm{d}$ map also likely overestimates the spatial extent of the protostellar heating relative to what would be needed for a $18.2''$ column density map, considering how compact the protostellar-heated sources are in the $160$ \textmu m data (see Figure \ref{fig:emission_maps}). Given that the \textit{Herschel} column density map was derived from a single-component spectral energy distribution (SED) model, such an error in the dust temperature estimation can propagate into the $18.2''$-resolution column density map. Indeed, Figure \ref{fig:emission_maps} shows halo-like artefacts around protostellar sources SVS 13 and IRAS 2 in the column density map. Considering that protostellar heating can also give rise to varying temperature profiles along the lines of sight, a single-component, isothermal SED model used to derive the column density map may fail to capture the observed emission adequately even if the dust temperature resolution is not a problem. For example, Appendix \ref{appendix:ColDen} shows that the $18.2''$- and $36''$ resolution column density gradients tend to agree well with each other, suggesting that any issues with the resolution are highly localized. Due to column density gradients being sensitive to these modelling limitations and the fact that dust emission and polarization are inherently temperature-weighted along the lines of sight, we caution against over-interpreting the $\nabla N(\mathrm{H}_2$) gradient results.

Comparable analyses with \textit{Planck} \citep[e.g.,][]{Planck2016_HROMagClouds} probe structure alignments using column density structures and on much larger scales.  As a result, most of their independent measurements are dominated by what is considered low-column density ($< 2.5 \times 10^{22}$ cm$^{-2}$) material in our study. Indeed, the fact that the $10'$ Planck beam is about 33 times that of our column density counterpart map suggests that their beams tend to sample more diffuse gas that is less shielded from the interstellar radiation fields (ISRF), even when pointed towards the highest column density regions. The \textit{Planck} column density map, derived from \textit{Planck} emission maps, is thus more biased towards the warmer gas than our \textit{Herschel}-derived column density map and is likely in better agreement with the gas traced by our dust emission maps. This similarity likely explains why our emission structural alignments with the $B$ field agree more with the Planck result than our column density alignments. We caution, however, that the scales of the \emph{Planck} study are substantially larger than those we study. For reference, the entirety of our BISTRO data of NGC 1333 is only about two of \emph{Planck}'s $10'$ beams in size. 

\cite{Soler2019} also performed similar analyses using the \textit{Planck} $B$ field measurements with $36''$ \textit{Herschel}-derived column density maps. In Perseus South, which contains NGC 1333, they found no preferential alignment ($Z_{x} \sim 0$) at $N(\mathrm{H}) \gtrsim 3 \times 10^{22}$ cm$^{-2}$. For the nearby ($d < 450$ pc) molecular clouds in general, they only found weak perpendicular alignments ($|Z_{x}| \lesssim 2\sigma$) at $N(\mathrm{H}) \gtrsim 8 \times 10^{21}$ cm$^{-2}$. While these results are consistent with ours, the scale for which the \textit{Planck} data probes $B$ field is still substantially larger than that probed by POL-2. It is thus difficult to interpret the significance of the non-alignment results in their study, both in relation to ours and those by \cite{Planck2016_HROMagClouds}. 

\subsection{Local Mass Flows in a Compressed Region}
\label{subsec:postshock_flow}

From the VCSs identified in NGC 1333, most of them showed no preferred alignment between the local velocity gradients and magnetic field. Only one region, \textit{VCS-e}, had significant alignment.  \textit{VCS-e} shows a preferential parallel alignment between its velocity gradient and the magnetic field.  This region is located on a ridge in NGC 1333 that \citetalias{Dhabal2019} interpreted to have been formed by large-scale compression, likely driven by an expanding bubble. \citetalias{Dhabal2019} argued that the qualitative alignment between their velocity gradients and the magnetic field measured by \cite{Doi2020} indicates the presence of mass flow along a magnetic field, similar to those found in shock-compressed regions in MHD simulations  \citep{ChenCY2014,ChenCY2015}. \citetalias{Dhabal2019}, however, only performed a single-velocity fit to their $4''$ NH$_3$ data and treated their entire $v_\mathrm{LSR}$ map of NGC 1333 as a single coherent region. We will thus be revisiting this region with the VCSs identified from the two-component fits to the 14.1$\arcsec$ NH$_3$ data.

The extended velocity front that \citetalias{Dhabal2019} identified as a source of compression actually corresponds to a fairly sharp $v_\mathrm{LSR}$ jump between two distinct groups of VCSs: the first group consisting of \textit{VCS-d} and \textit{VCS-e} with lower $v_\mathrm{LSR}$ values of $\sim 7$ km s$^{-1}$ and a second, adjacent group consisting of \textit{VCS-a} and \textit{VCS-b} with higher $v_\mathrm{LSR}$ values of $\sim 8$ km s$^{-1}$  (see Fig. \ref{fig:kinMaps} \& Fig. \ref{fig:regions}). The velocity gradient front inferred by \citetalias{Dhabal2019}, therefore, may not necessarily correspond to a kinematically continuous structure (i.e., velocity coherent), and should be considered inter-VCS in nature. The velocity gradients we measured, however, are strictly within the identified VCSs, and thus, they are intra-VCS by definition. 

Despite our velocity gradients being intra-VCSs, the $\nabla v$ we found in \textit{VCS-e} still aligns perpendicularly to the magnetic field, similar to the inter-VCSs $\nabla v$ found by \citetalias{Dhabal2019} between \textit{VCS-a} and \textit{VCS-e}. Such an agreement suggests that a common process is driving these two types of velocity gradients. For example, the inter-VCS and intra-VCS velocity may trace the global and local convergence flows seen in the MHD simulations by \citet{ChenCY2014} and \citet{ChenCY2015}. In these simulations, the global flow is driven by large-scale turbulence, and the local flow is driven by gravity. Under this scenario, the initial turbulent compression first forms a denser post-shock region. The overdensities (e.g., filaments) that emerged within this region then subsequently drive accretion along the $B$ field via self-gravity. Such an accretion can manifest in the preferential parallel alignments between the $B$ field and $\nabla v$, like those we found in \textit{VCS-e}. Indeed, synthetic observations of these simulations, focused on a similar density regime as our observations, were compared with Serpens South observations to infer such flows \citep{ChenCY2020}.

To quantify the ratio between the gravitational potential and the kinetic energies of the flow observed in \textit{VCS-e}, we adopt the analytic expression $C_v \equiv \Delta v_\mathrm{h}^2 / (G M/L )$ proposed by \cite{ChenCY2020}. Here, $\Delta v_\mathrm{h}$ is the LOS velocity difference across half a filament, $M/L$ is the filament's mass per unit length, and $G$ is the gravitational constant. A $C_v$ value less than unity indicates a flow driven predominately by gravity, while a $C_v$ greater than unity indicates a flow driven by non-gravitational processes such as turbulence or expanding bubbles. For \textit{VCS-e}, we measure an intra-VCS $\Delta v_\mathrm{h}$ of $\sim 0.3$ km s$^{-1}$. If we adopt $M/L = 83$ M$_\odot$ pc$^{-1}$ from \cite{Hacar2017} for their \textit{filament 10}, the structure onto which \textit{VCS-e} seemingly accretes, then resulting ratio value is $C_v \sim 0.3$. This $C_v$ value is well below unity and indicates that the flow within \textit{VCS-e} is gravitationally driven and is consistent with those found with synthetic observations of simulations by \cite{ChenCY2020} at the epoch when filaments have formed prominently.

Since gas flows physically moving towards each other can manifest as distinct VCSs in observations \citep{Clarke2018}, the inter-VCS velocities can still contain important information regarding converging flows, such as those driven by expanding bubbles. If we instead use the inter-VCS $\Delta v_\mathrm{h}$ of $\sim 1.0$ km s$^{-1}$, a lower limit measured between \textit{VCS-e} and \textit{VCS-a} in the region centred on the \textit{filament 10} from \cite{Hacar2017}, then we get $C_v \sim 3.1$, which is significantly larger than unity. Similarly, \citetalias{Dhabal2019} found $C_v > 1$ at and around \textit{VCS-e}. Such a $C_v$ would indicate a flow that is driven predominately by non-gravitation means.

The discrepancy between the intra- and inter-VCS $C_v$ values, however, are not necessarily contradictory. In fact, these two values can be reconciled under a single theoretical model, where the former and the latter correspond to the two stages of dense filament formation in simulations by \cite{ChenCY2014, ChenCY2015} and \citeauthor{Inoue2018} (\citeyear{Inoue2018}; see also review by \citealt{Pineda2023}). Under this scenario, a large-scale compressing flow, driven by turbulence or an expanding bubble, can produce distinct observable VCSs with inter-VCS $C_v$ values larger than unity. When such a flow collides and becomes relatively stagnant in the compressed region, the overdensities within the region can drive gravity-dominated accretion flows, resulting in velocity-continuous intra-VCS flows with $C_v < 1$, such as in \textit{VCS-e}. The simultaneous observations of these two flows suggest that the initial turbulent flows may have only made partial contact and, thus, not completely destroyed in the collisional process forming the compressed region. Two-component spectral fitting and VCS identification, however, are needed to identify these flows properly.

\subsection{Core Contraction or Rotation}
\label{subsec:vgrad_contraction}

Even though \textit{VCS-h}, the largest VCS in NGC 1333 by spatial extent, does not show a preferred alignment between $B$ and $\nabla v$ overall, the two fields are preferentially aligned perpendicularly at pixels with $N(\mathrm{H}_2) \geq 5\times 10^{22}$ cm$^{-2}$ ($> 2.5\sigma$ detection; see Fig. \ref{fig:HRO_vcs}).  These high column density regions are fairly compact and likely dominated by the dense cores embedded in the filament \citep[e.g.,][]{Doi2020}.  As such, the perpendicular alignments between $B$ and $\nabla v$ may indicate gravitational contraction towards dense cores along a direction that is perpendicular to the $B$ field support rather than flowing along the field lines. Alternatively, the perpendicular alignment could be due to the magnetic field and the velocity gradient tracing different scales in gravitationally dominant regions. Under this scenario, the contraction flows traced by the velocity gradient can significantly distort the magnetic field inside dense structures and consequently depolarize the dust emission there \citep[e.g.,][]{ChenCY2016}. This causes the observed dust polarization to predominately trace the natal magnetic field outside the dense region that is perpendicular to the gas flow that distorts the magnetic field inside \citep[e.g.,][]{Gomez2018}.

Rather than gravitational contraction, the $\nabla v$ seen toward these dense regions could be dominated by core rotation instead. In this case, the preferential perpendicular alignment between $B$ and $\nabla v$ would be contrary to the results found in simulations by \cite{ChenCY2018} and the 10\arcmin \ large-scale study conducted with {\em Planck} polarization and single-dish $32''$ NH$_3$ data by \cite{Pandhi2023}. Those studies both found no alignment between core angular momentum and the global $B$ field. The number of compact, dense structures in \textit{VCS-h} adopted by \citet{Doi2020}, however, is low (n $\sim 5$). Moreover, the {\em Planck} data used by \cite{Pandhi2023} has a substantially lower resolution than our $B$ field data and thus traces much larger-scale $B$ fields. Further comparison between our result and their work must account for these differences. More RO measurements that extend throughout NGC 1333 with more sensitive $B$ field measurements and better NH$_3$ spatial coverage will also need to be investigated further. 

\subsection{Correlations with Gas Properties}
\label{subsec:gas_properties}

In Section \ref{subsub:local_sgrad}, we showed that PRS of individual local regions does not correlate with their respective median gas properties. In this section, we examine the PRS in different gas property regimes sampled globally based on their percentile values rather than their spatial proximity. Specifically, we divide our samples into equal-size bins by percentile ranges for $N(\mathrm{H}_2)$, $T_\mathrm{d}$, $|\nabla v|$, and $\sigma_v$ and compute a $Z_\mathrm{x}$ value for each bin, with percentile values computed only for pixels where the gas property and the relative orientation are robustly measured. This approach tests the general field alignment for material under similar physical conditions within NGC 1333 regardless of where these pixels are found. Using this approach, we found trends between $Z_{BI}$ and these four gas properties, but not for $Z_{BN}$ and $Z_{Bv}$. We will thus focus solely on $Z_{BI}$ in our discussion here.

\begin{figure}
    \includegraphics[width=\columnwidth]{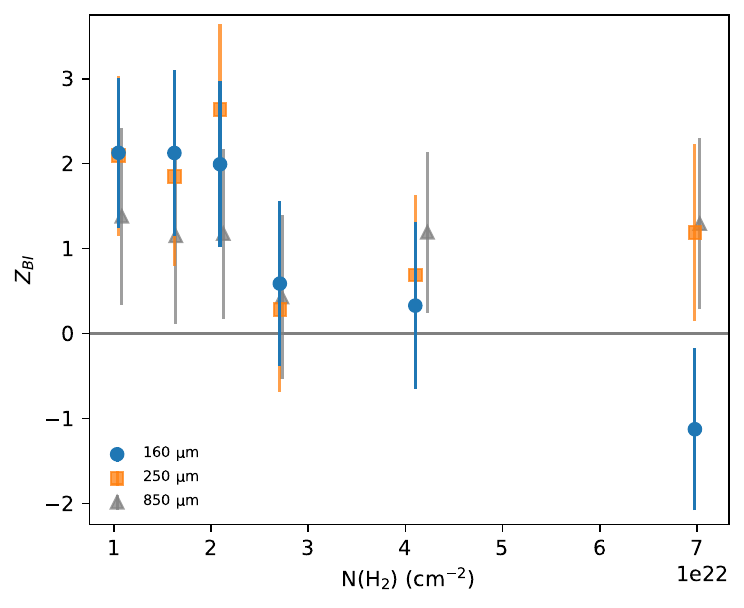}
    \caption{PRS values between $B$ and the 160, 250, and 850 \textmu m emission maps (i.e., $Z_{BI}$) measured over N(H$_2$) ranges defined by equal-size bins and represented by the median bin value. The size of each bin is 70 pixels. The different symbols correspond to each of the emission maps, and the error bars show 1 $\sigma$ uncertainties.} %
    \label{fig:Z_vs_Colden}
\end{figure}

\subsubsection{Alignment Correlation with Column Densities}
\label{subsub:Z_column_den}
We divided NGC 1333 into six equal-sized bins of 70 pixels based on column density ranges in increments of 16.7-percentiles, and then measured their corresponding PRS. We only use independent RO measurements for the PRS calculation. We further note that the RO measurements in these column density bins are from pixels that are located spatially throughout the cloud, making many of our samples spatially independent beyond just the beam separation. 

Figure \ref{fig:Z_vs_Colden} shows the PRS for each of the emission gradients, $Z_{BI}$, plotted against their respective median bin $N(\mathrm{H}_2)$ values. The 160 \textmu m data show a general anti-correlation for $Z_{BI}$, similar to that found in Oph A on a comparable scale using 154 \textmu m observations \citep{LeeD2021}.  Such a trend indicates that the magnetic field starts out being parallel to the dust emission gradients (perpendicular to elongated structures) at our lowest column densities ($\sim 1 \times 10^{22}$ cm$^{-2}$) but then transitions to no preferred alignment or more perpendicular alignment at higher column densities. For the 250 \textmu m and 850 \textmu m emission maps there are hints of an anticorrelation for $N(\mathrm{H}_2) \lesssim 4.5 \times 10^{22}$ cm$^{-2}$ too, but the trend is less clear.

Higher (more positive) $Z_{BI}$ values in the lowest column density bins are consistent with the results of \cite{Planck2016_HROMagClouds}.  Our parallel alignment between $B$ and the emission gradients match the general \emph{Planck}-derived perpendicular alignment to the structure elongations at $N(\mathrm{H}) \gtrsim 5 \times 10^{21}$ cm$^{-2}$, although this transitional column density is well below all our samples in NGC 1333. Even though we probe spatial scales that are much smaller than those from the \emph{Planck} study, synthetic observations of MHD simulations on our scales show that such an alignment can still arise from accretion along magnetic fields when gravity starts to dominate over magnetic support \citep{ChenCY2016}. Indeed, considering all the dense filaments in NGC 1333 are significantly thermally supercritical (e.g., \citealt{Hacar2017}; Chen et al. 2023 submitted), the parallel alignment between the emission gradient and $B$ in our lower column density bins likely results from gravity dominating over magnetic support.

As gravity becomes more dominant towards higher densities, contracting structures can drag and distort the magnetic field significantly inward. The complexity of such a distortion in 3D can cause the projected 2D alignments to shift or wash out, resulting in depolarization \citep{ChenCY2016}. Indeed, \cite{Doi2021} found filament widths measured in polarized emission in NGC 1333 to be consistent with pinched magnetic field distortion under simple contraction models. Unresolved complex field morphologies likely explain why our $Z_{BI}$ trends toward no preferred alignment with increasing column densities.  

Similarly, field alignment can also change at the highest column densities due to the $B$ field being dragged along and becoming relatively parallel to elongated structures like filaments, similar to those seen in simulations by \citep{Gomez2018}. Observationally, \citet{Pillai2020} found a transition to parallel structure alignment at $N$(H$_2) \gtrsim 2 \times 10^{22}$ cm$^{-2}$ in Serpens South's filament-hub system using 214 \textmu m polarization data. \citet{Monsch2018} also saw parallel alignment between $B$ and a high-density filament in Orion A using 850 \textmu m polarization data.  Therefore, the extended trend of the 160 \textmu m data having $B$ transition to perpendicular alignment relative to the emission gradient (parallel to elongated structures) in Figure \ref{fig:Z_vs_Colden} may similarly correspond to field lines being dragged by gravitational distortions. We caution, however, that the PRS statistics in the highest column density bin have weak alignment signals ($|Z_{BI}| \sim 1$) and more data are needed to confirm whether this trend is robust and unique to the 160 \textmu m emission structures.  

\begin{figure}
    \includegraphics[width=\columnwidth]{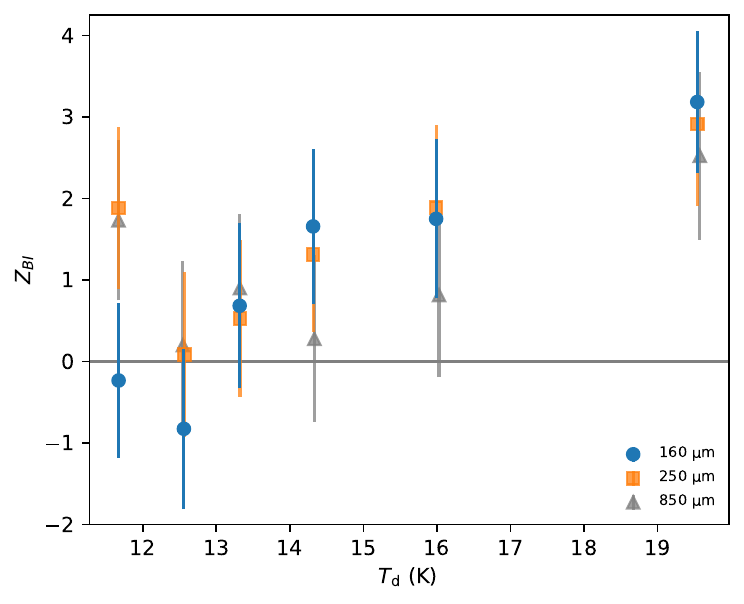}
    \caption{Same as Figure \ref{fig:Z_vs_Colden}, but calculated from 70-pixel-size $T_\mathrm{d}$ bins.}
    \label{fig:Z_vs_Td}
\end{figure}

\subsubsection{Alignment Correlation with Dust Temperatures}
\label{subsub:Z_dust_temp}

Figure \ref{fig:Z_vs_Td} shows $Z_{BI}$ plotted against their respective $T_\mathrm{d}$ values for six temperature-binned regimes using 70-pixel bins (following Section \ref{subsub:Z_column_den}). In general, $Z_{BI}$ shows a positive correlation between $Z_{BI}$ and temperature for all three dust emission maps.  A positive correlation with the temperature suggests that the $B$ field tends to align parallel to the emission gradients (perpendicular to the structure) in warmer gas.  Since our high-temperature pixels are dominated by diffuse, larger-scale gas around the B-stars (see Figure \ref{fig:emission_maps}), the less-shielded diffuse gas structures appear to be most aligned with the magnetic field. Indeed, the highest $T_\mathrm{d}$ samples seen in Figure \ref{fig:Z_vs_Td} are mostly found at \textit{Region 13}, which is predominately heated by B stars in their vicinity (\citealt{ChenMike2016}). This warm watershed region's PRS also showed the strongest parallel alignment with the emission gradient in NGC 1333 (see Section \ref{subsub:local_sgrad}). Assuming denser gas is better shielded and hence colder than its more diffuse counterparts, the correlation between $Z_{BI}$ and $T_\mathrm{d}$ is also consistent with the anti-correlation between $Z_{BI}$ and $N(\mathrm{H}_2)$ seen in Figure \ref{fig:Z_vs_Colden}.

Similar to the highest column density $Z_{BI}$ values seen in Figure \ref{fig:Z_vs_Colden}, the intensity gradients only deviate from the same general trend at the lowest $T_\mathrm{d}$ bin. Since the highest $N$(H$_2$) and the lowest $T_\mathrm{d}$ bins likely trace the densest pre- and protostellar cores, the deviation in the trend may be associated with early instances of localized collapse. The lack of elevated $T_\mathrm{d}$ near the protostellar sample in the lowest $T_\mathrm{d}$ bin suggests that their associated YSOs are fairly young and deeply embedded. The different behaviour of $Z_{BI}$ between the 160 \textmu m data and the 250 \textmu m and 850 \textmu m data in these bins are likely due to 160 \textmu m map being more sensitivity to the warmer gas along lines of sight, e.g., the YSO-heated inner region of a protostellar core or the warmer envelopes of a prestellar core.

\begin{figure}
    \includegraphics[width=\columnwidth]{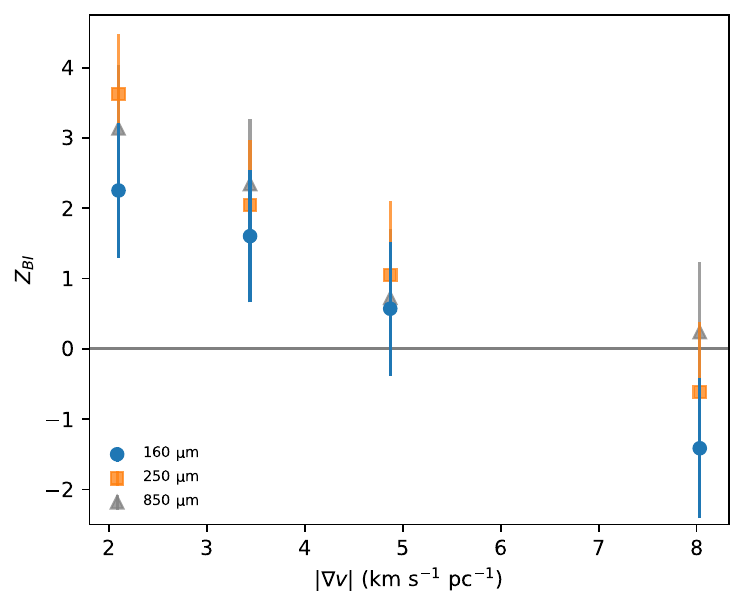}
    \caption{Same as Figure \ref{fig:Z_vs_Colden}, but calculated from 50-pixel-size $|\nabla v|$ bins.}
    \label{fig:Z_vs_Vgrad}
\end{figure}

\subsubsection{Alignment Correlation with Gas Kinematics}
\label{subsub:Z_vgrad}

To see if the magnetic field shows any preferred alignment with the gas kinematics, we calculated $Z_{BI}$ with pixels binned according to their $|\nabla v|$ and $\sigma_v$ values. We use 50-pixel bins for both measurements, corresponding to 25-percentile increments (following the same approach as with the column density maps in Section \ref{subsub:Z_column_den}). We chose a smaller bin size here to accommodate the smaller overlaps between the $B$ field and the NH$_3$ detections and ensure at least four PRS measurements across each gas property.

\begin{figure}
    \includegraphics[width=\columnwidth]{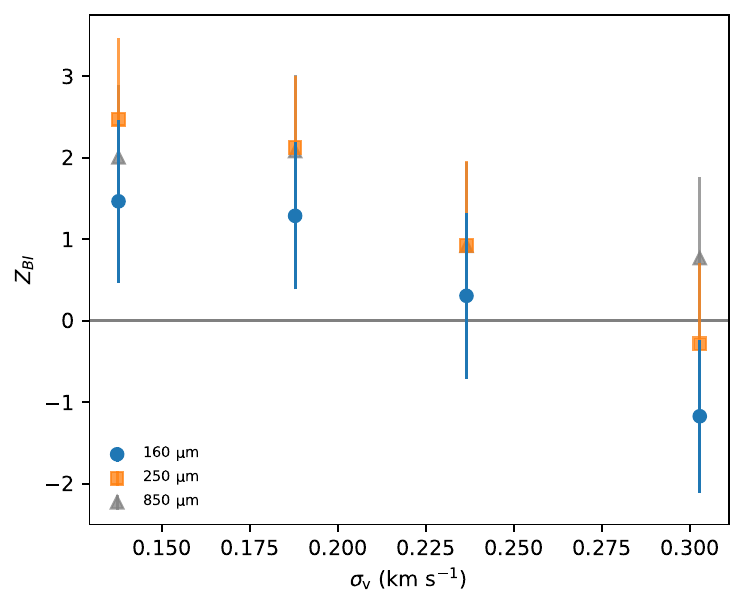}
    \caption{Same as Figure \ref{fig:Z_vs_Colden}, but calculated from 50-pixel-size $\sigma_v$ bins.}
    \label{fig:Z_vs_Sigv}
\end{figure}

Figure \ref{fig:Z_vs_Vgrad} and \ref{fig:Z_vs_Sigv} show trends of $Z_{BI}$ in relation to $|\nabla v|$ and $\sigma_v$, respectively, for each emission map. We find an anti-correlation between $Z_{BI}$ and both $|\nabla v|$ and $\sigma_v$ in all three wavelengths. The similarity between the two anti-correlations, however, is unexpected, given that $|\nabla v|$ and $\sigma_v$ exhibit very different morphologies and do not correlate well with each other (e.g., see Fig. \ref{fig:kinMaps}). While velocity gradients can arise from large-scale flows, rotation, or contraction (see Section \ref{subsec:vgrad_contraction}), the velocity gradient in NGC 1333 on the scale we probe ($\sim 0.02$ pc) seems to be dominated by ripple structures (see Figure \ref{fig:kinMaps}). These ripples appear to be wave-like in nature, likely driven by gravo-acoustic or MHD waves. The elevated $\sigma_v$, on the other hand, is most likely turbulent in nature, given that the NH$_3$ gas temperatures do not vary significantly in NGC 1333 \citep[][]{GAS}. These elevated $\sigma_v$ can be driven by processes such as accretion. We further discuss the relationship between $Z_{BI}$ and $|\nabla v|$, as well as $\sigma_v$, in the next section (\ref{subssec:Z_vgrad}) and Section \ref{subsec:Z_sigv_infalls}, respectively.

\subsection{Disturbance from MHD Waves}
\label{subssec:Z_vgrad}

The ripple-like features in the $|\nabla v|$ maps (see Fig. \ref{fig:kinMaps}) have been previously seen with NH$_3$ in NGC 1333 on the 0.05 pc scale \citep{ChenMike2020} and in Perseus B5 on the 0.01 pc scale \citep{ChenMike2022}. These features appear to be small-scale perturbations on top of the larger-scale velocity gradients that likely trace mass transports such as accretion flows \citep[e.g.,][]{ChenCY2020}. When taking a few sample slices through the ripples perpendicularly, we find the ridge-to-ridge and valley-to-valley distance in our data to be quasi-periodic, with lengths that range between 0.02 pc to 0.04 pc. 

Similar ripples have been found along filaments in line-of-sight velocities observed with other tracers, such as CO (e.g., \citealt{Hacar2013}) and N$_2$H$^+$ (e.g., \citealt{Tafalla2015}), as well as in hydrodynamic simulations (e.g., \citealt{Clarke2016}). In their simulations, \citet{Clarke2016} found ripples in supercritical filaments seeded initially by gravo-acoustic oscillations in the subcritical accretion phase of the filament. In magnetohydrodynamic (MHD) models, Alfv\'{e}n and magnetosonic waves are able to propagate in directions parallel and perpendicular to the unperturbed magnetic fields, respectively \citep[e.g.,][]{Tritsis2016,Tritsis2018}. Such waves can be excited by stellar feedback, such as outflows and expanding stellar wind shells, and can enhance turbulent energy transfer from the local scale to the cloud scale \citep{Offner2018}. Given that NGC 1333 is a fairly active star-forming region with a large number of outflows (e.g., \citealt{Arce2010}), we expect it to be capable of exciting such waves. 

As seen in Figure \ref{fig:field_maps}d, the magnetic field tends to align perpendicular to the ripple features in many places, particularly toward \textit{Regions 9} and \textit{10}. If we assume valleys of the $|\nabla v|$ ripples trace wavefronts, where wave motions are expected to be locally stagnant, then our results would be consistent with the expected behaviour of Alfv\'{e}n waves, where the waves propagate along the $B$ field. In some limited areas, however, such as the southeastern part of \textit{Region 11}, the ripples can align in parallel with the magnetic field, consistent with the behaviour of magnetosonic waves. Since magnetosonic waves can be non-linearly coupled with Alfv\'{e}n waves \citep{Heyvaerts1983}, it is possible that both types of waves remain important in NGC 1333. We note that inclination effects must be properly accounted for to interpret the wave propagation directions well. Nevertheless, the preferential alignments between the prominent $|\nabla v|$ ripples and the magnetic field suggest that these ripples are driven by MHD waves.

Regardless of the exact nature of the $|\nabla v|$ ripples, pixels with lower $|\nabla v|$ values likely trace the gas least perturbed by these ripples (see Figure \ref{fig:Z_vs_Vgrad}). The strong parallel alignment between the $B$ field and emission gradients at the lower $|\nabla v|$ bin may thus correspond to a ``pristine'' alignment, similar to those we found globally in NGC 1333 (see Figure \ref{fig:global_histo}). Such a global alignment is dominated in pixel count by the warmer, more diffuse gas with emission gradients aligned preferentially parallel to the magnetic field (see Section \ref{subsub:Z_column_den} and \ref{subsub:Z_dust_temp}). If such a parallel gradient alignment (i.e., perpendicular structure alignment) is indeed pristine, then it is likely inherited from the same trend found on the larger scales for $N(\mathrm{H}) \geq 5 \times 10^{21}$ cm$^{-2}$ \citep{Planck2016_HROMagClouds}, which corresponds to the initial configuration in this density regime before gravity distorts the $B$ field significantly on the smaller scales (see Section \ref{subsub:Z_column_den}). 

The anti-correlation between $Z_{BI}$ and $|\nabla v|$ seen in Figure \ref{fig:Z_vs_Vgrad} thus suggests that as $|\nabla v|$ increases, the $B - \nabla v$ alignment becomes more perturbed from such pristine alignments. Given that most of the higher $|\nabla v|$ pixels come from ripple-like structures, MHD waves are likely the primary driver of these perturbations. Nevertheless, since many of the Class 0/I YSOs identified by \cite{Dunham2015} are located within a beam of the highest $|\nabla v|$ bin pixels, instances of local collapse may also contribute to the trends in that bin. Indeed, the $|\nabla v|$ structure indicative of an infall towards the densest prestellar ``condensate'' in Persues B5 have distinctly higher $|\nabla v|$ values than the ripple structures seen throughout the B5 sub filaments \citep{ChenMike2022}. 

\subsection{Disturbances from Infalls}
\label{subsec:Z_sigv_infalls}

Similar to the $Z_{BI} - |\nabla v|$, we find an anticorrelation between $Z_{BI}$ and $\sigma_v$ (see Fig. \ref{fig:Z_vs_Sigv}). This anticorrelation, however, is unlikely to be driven by the decrease in $Z_{BI}$ with increasing $|\nabla v|$ discussed in the previous section. The fact that $\sigma_v$ and $|\nabla v|$ do not correlate well with each other and are morphologically dissimilar (see Figure \ref{fig:kinMaps}) implies that the two trends are not causally linked, at least not directly. There are a few exceptions where compact areas with $\sigma_v > 0.5$ km s$^{-1}$ appear to correlate with high $|\nabla v|$ regions, but these areas are mostly located outside of where the $B$ field is robustly detected (see Fig. \ref{fig:field_maps}).

Interestingly, despite protostellar cores having larger $\sigma_v$ than their prestellar counterparts when observed at a lower resolution \citep{KirkHelen2007}, the pixels neighbouring Class 0/I YSOs in our VCSs (within $\sim 1$ beam) do not necessarily display significantly higher $\sigma_v$ values compared to each VCS's overall $\sigma_v$. The median $\sigma_v$ within a beam of Class 0/I YSOs is 0.16 km s$^{-1}$ while the median $\sigma_v$ within the VCSs is 0.22 km s$^{-1}$. This lack of elevated $\sigma_v$ near YSOs suggests that accretion-driven turbulence associated with protostellar infall (e.g., \citealt{Klessen2010}) is not the main driver behind the elevated NH$_3$ $\sigma_v$ seen in VCSs, particular where the $B$ field is robustly detected. 

We also find no correlation between elevated $\sigma_v$ and locations of protostellar outflows, indicating that higher $\sigma_v$ regions in NGC 1333 are not driven by outflows either, consistent with the single-dish NH$_3$ findings in NGC 1333 \citep{ChenMike2020}. Instead, higher $\sigma_v$ pixels tend to be found towards the outskirts of VCSs. This behaviour is qualitatively similar to the ``transitions to coherence'' towards denser structures found by single-dish observations (e.g., \citealt{Pineda2010}) and the trend found by \cite{ChenMike2022} at a higher resolution, where the NH$_3$ $\sigma_v$ increases radially from subfilament spines. Such a trend can result from the top-down turbulence cascades from the cloud \citep{Federrath2016}, a consequence of ongoing hierarchical collapse \citep{Vazquez-Semadeni2019} or accretion-driven turbulence \citep{Heigl2020}. Due to limited spatial overlap between $B$ field detections and the VCS footprints, our RO measurements tend to be found in the lower $\sigma_v$ regions towards denser structures, further from the VCS edges (see Fig. \ref{fig:field_maps}). The only exception is the northern part of \textit{VCS-h} around \textit{Region 11} (see Fig. \ref{fig:regions}), where the $B$ field is detected towards lower column density structures due to higher dust temperatures in the region.

The anti-correlation between $Z_{BI}$ and $\sigma_v$ indicates that the physical mechanism that drives increased $\sigma_v$ can also perturb the alignments between $B$ and the emission gradients from its pristine parallel state. As $\sigma_v$ increases, such alignments become less parallel (see Fig. \ref{fig:Z_vs_Sigv}). The perturbation behind this trend may predominately come from disturbances to the $B$ field due to accretion, which can result in increased $\sigma_v$ in our observations either directly as unresolved infall motions \citep{Vazquez-Semadeni2019} or indirectly as accretion-driven turbulence \citep{Heigl2020}. While the top-down turbulent cascade \citep{Federrath2016} can also explain the behaviour of increased $\sigma_v$ in NGC 1333, if the $B$ field alignment was disturbed purely by cascading turbulence rather than local infall, then the effect would be more pronounced in the larger-scale diffuse gas than their denser counterparts. Our finding where $Z_{BI}$ anticorrelates with column density (see Section \ref{subsub:Z_column_den}), however, does not support this interpretation. The decrease of $Z_{BI}$ towards higher $\sigma_v$ is thus more likely caused by accretion-driven disturbance to the alignments than by the top-down turbulence cascade.

\section{Conclusions}
\label{sec:conclusions}

For the NGC 1333 region, we examined the alignment between the local magnetic field, $B$, inferred from POL-2/JCMT observations and the gradient fields from maps of thermal dust emission at 850 \textmu m, 250 \textmu m, and 160 \textmu m, \emph{Herschel}-derived column density, and NH$_3$ velocity centroid, i.e., $\nabla I_{850}$,  $\nabla I_{250}$, $\nabla I_{160}$, $\nabla N(\mathrm{H}_2)$, and $\nabla v$, respectively.  For the velocity gradients, we first fit two-component models to the VLA+GBT NH$_3$ observations using the \texttt{MUFASA} software \citep{ChenMike2022} and then identified velocity-coherent structures (VCSs) in position-position-velocity space using the \texttt{DBSCAN} clustering algorithm \citep{Ester1996}. We then take the gradients of each VCSs' velocity centroid maps to obtain $\nabla v$.

We calculate the relative orientation (RO) angles, $\phi$, between the $B$ field and each gradient field under the convention where $0^\circ$ and $90^\circ$ represent parallel and perpendicular alignments, respectively.  We then quantified the distribution of these ROs in local gas structures and examined how they correlate with different gas properties using the Projected Rayleigh Statistics (PRS). Here, we summarize our main results:

\begin{enumerate}
  \item Globally, we find a preferential parallel alignment between the $B$ field and each of the emission gradients (i.e., $\nabla I_{160}$, $\nabla I_{250}$, and $\nabla I_{850}$) with RPS $Z_\mathrm{x} > 2.5\sigma$. Since emission gradients are orthogonal to structure elongations, this result indicates that gas structures in NGC 1333 tend to align perpendicular to the $B$ field on average, similar to the results found for dense, elongated cloud structures on the larger scales \citep{Planck2016_HROMagClouds}. The $B$ field and the $\nabla v$ and $\nabla N$(H$_2$) fields, however, do not show a preferred alignment globally.
  
  \item Only a few VCSs showed preferred alignment between $B$ and $\nabla v$ locally. \textit{VCS-e} showed preferential parallel alignment while \textit{VCS-h} showed preferential perpendicular alignment at higher densities. The parallel alignment in \textit{VCS-e} can be explained by a bubble compression front \citep{Dhabal2019}. The gravo-kinematic energy ratio \citep[i.e., $C_v$;][]{ChenCY2020} of \textit{VCS-e} indicates that its internal gas flow is driven predominantly by gravity while the inter-VCS velocity differential suggests that \textit{VCS-e} is being pushed into other VCSs by non-gravitational means or vice versa, presumably by the expanding bubble. These two behaviours are consistent with the two stages of filament and core formation via MHD gas compression under a single scenario \citep{ChenCY2014, ChenCY2015}. The perpendicular alignment in \textit{VCS-h} at high densities, on the other hand, could be tracing areas of local infall.
  
  \item Most subregions defined by 850 \textmu m emission do not show significant alignments between the $B$ field and the emission gradients. \textit{Region-9} and \textit{13}, however, have preferential parallel alignment while \textit{Region-14} has preferential perpendicular alignment. \textit{Region 13}, in particular, is located in the warm diffuse region heated by the two nearby B stars (see \citealt{ChenMike2016}).
  
  \item Broadly, the RO distributions measured for each of the defined subregions show no correlation with the local properties of these regions, such as their median $N$(H$_2$) or $T_\mathrm{d}$ values. This lack of correlation indicates that the local variations in alignments within the region wash each other out over the different properties.
  
  \item When binned according to gas properties across NGC 1333, rather than their spatial proximity, we found $Z_\mathrm{x}$ calculated for the ROs between the $B$ field and the emission gradients (i.e., $Z_\mathrm{BI}$) correlates with $T_\mathrm{d}$ and anti-correlates with N(H$_2$), $|\nabla v|$, and $\sigma_v$. The correlation with $T_d$ and anti-correlation with N(H$_2$) indicates that the $B$ field aligns more in parallel to the emission gradients (perpendicular to structures) in warmer, more diffuse gas. Such alignments become more distorted, likely due to gravity, as the gas gets colder and denser. The anticorrelation between $Z_\mathrm{BI} - |\nabla v|$ and $Z_\mathrm{BI} - \sigma_v$ suggests that the $B$ field in NGC 1333 aligns more perpendicular to structures in less kinematically perturbed gas. These less perturbed alignments are similar to the global average we found for NGC 1333 and that found on larger scales \citep{Planck2016_HROMagClouds}. 

  \item The $|\nabla v|$ within VCSs shows ripple-like structures that are spaced about 0.02 pc to 0.04 pc apart (peak-to-peak and trough-trough). These ripples can be coherently aligned perpendicular or parallel to the $B$ field in several regions, indicating they may correspond to Alfv\'{e}n and magnetosonic waves propagating parallel and perpendicular to the $B$ field, respectively (e.g., \citealt{Tritsis2016}).

\end{enumerate}

Our results here establish that the magnetic field's relationship with gas structures and mass flows changes on smaller (local) scales. Pixels associated with warmer, lower column density gas dominate our sample by number and generally agree with the findings of \cite{Planck2016_HROMagClouds} in the same density regime, where the magnetic field aligns perpendicularly to the gas structures. The lower-density gas in our study may, therefore, represent the ``pristine'' state inherited from the larger scales. The disturbances from this state are limited to localized areas as the gas becomes denser and colder, where perturbations can arise from processes such as infall, accretions, and MHD waves.

Sensitive polarization data and synthetic observations of simulations that can capture large and small scales will be crucial to interpret our results robustly. Such studies will be particularly valuable to understand the role of temperature weighting along the lines of sight, 
how the observed $B$ field co-evolves with dense structures, and which physical processes can produce the trends we found. Furthermore, investigating $|\nabla v|$ in synthetic observations will help understand the origin of the $|\nabla v|$ ripples, how they interact with magnetic fields, and whether or not they are driven by MHD waves. 

\section*{Acknowledgements}

The authors thank Che-Yu Chen for excellent and insightful discussions on the interpretation of our results. S.I.S acknowledge support from NSERC Discovery Grants RGPIN-2020-03981, RGPIN-2020-03982, RGPIN-2020-03983 and a Research Initiation Grant from Queen’s University. L.M.F. acknowledges support from an NSERC Discovery Grant (RGPIN-2020-06266) and a Research Initiation Grant from Queen’s University. D.J. is supported by NRC Canada and by an NSERC Discovery Grant. K.P. is a Royal Society University Research Fellow, supported by grant number URF$\backslash$R1$\backslash$211322. M.Tamura is supported by JSPS KAKENHI grant No.18H05442. J.Kwon is supported by JSPS KAKENHI grant No.19K14775. M.Tahani is supported by the Banting Fellowship (Natural Sciences and Engineering Research Council Canada) hosted at Stanford University. The work of M.G.R is supported by NOIRLab, which is managed by the Association of Universities for Research in Astronomy (AURA) under a cooperative agreement with the National Science Foundation.

The James Clerk Maxwell Telescope is operated by the East Asian Observatory on behalf of The National Astronomical Observatory of Japan; Academia Sinica Institute of Astronomy and Astrophysics; the Korea Astronomy and Space Science Institute; the National Astronomical Research Institute of Thailand; Center for Astronomical Mega-Science (as well as the National Key R\&D Program of China with No. 2017YFA0402700). Additional funding support is provided by the Science and Technology Facilities Council of the United Kingdom and participating universities and organizations in the United Kingdom and Canada. Additional funds for the construction of SCUBA-2 and POL-2 were provided by the Canada Foundation for Innovation. The James Clerk Maxwell Telescope has historically been operated by the Joint Astronomy Centre on behalf of the Science and Technology Facilities Council of the United Kingdom, the National Research Council of Canada and the Netherlands Organisation for Scientific Research. The authors wish to recognize and acknowledge the very significant cultural role and reverence that the summit of Maunakea has always had within the indigenous Hawaiian community.  We are most fortunate to have the opportunity to conduct observations from this mountain.

This research has made use of data from the Herschel Gould Belt survey (HGBS) project (http://gouldbelt-herschel.cea.fr). The HGBS is a Herschel Key Programme jointly carried out by SPIRE Specialist Astronomy Group 3 (SAG 3), scientists of several institutes in the PACS Consortium (CEA Saclay, INAF-IFSI Rome and INAF-Arcetri, KU Leuven, MPIA Heidelberg), and scientists of the Herschel Science Center (HSC).


\section*{Data Availability}


The public accessibility of the data used in this work is listed below.

The NH$_3$ VLA data is available under project number 3A-309.

The GAS NH$_3$ data is available at:

\url{https://dataverse.harvard.edu/dataverse/GAS_DR1}.

The CO-subtracted JGBS 850 \textmu m data is available at:

\url{https://doi.org/10.11570/18.0005}

The HGBS data is available at:

\url{http://gouldbelt-herschel.cea.fr/archives}.



\bibliographystyle{mnras}
\bibliography{mnras_manuscript} 

\begin{thebibliography}{}
\makeatletter
\relax
\def\mn@urlcharsother{\let\do\@makeother \do\$\do\&\do\#\do\^\do\_\do\%\do\~}
\def\mn@doi{\begingroup\mn@urlcharsother \@ifnextchar [ {\mn@doi@}
  {\mn@doi@[]}}
\def\mn@doi@[#1]#2{\def\@tempa{#1}\ifx\@tempa\@empty \href
  {http://dx.doi.org/#2} {doi:#2}\else \href {http://dx.doi.org/#2} {#1}\fi
  \endgroup}
\def\mn@eprint#1#2{\mn@eprint@#1:#2::\@nil}
\def\mn@eprint@arXiv#1{\href {http://arxiv.org/abs/#1} {{\tt arXiv:#1}}}
\def\mn@eprint@dblp#1{\href {http://dblp.uni-trier.de/rec/bibtex/#1.xml}
  {dblp:#1}}
\def\mn@eprint@#1:#2:#3:#4\@nil{\def\@tempa {#1}\def\@tempb {#2}\def\@tempc
  {#3}\ifx \@tempc \@empty \let \@tempc \@tempb \let \@tempb \@tempa \fi \ifx
  \@tempb \@empty \def\@tempb {arXiv}\fi \@ifundefined
  {mn@eprint@\@tempb}{\@tempb:\@tempc}{\expandafter \expandafter \csname
  mn@eprint@\@tempb\endcsname \expandafter{\@tempc}}}

\bibitem[\protect\citeauthoryear{{Akaike}}{{Akaike}}{1974}]{Akaike1974}
{Akaike} H.,  1974, IEEE Transactions on Automatic Control, \href
  {http://adsabs.harvard.edu/abs/1974ITAC...19..716A} {19, 716}

\bibitem[\protect\citeauthoryear{{Andersson}, {Lazarian}  \&
  {Vaillancourt}}{{Andersson} et~al.}{2015}]{Andersson2015}
{Andersson} B.~G.,  {Lazarian} A.,   {Vaillancourt} J.~E.,  2015, \mn@doi
  [\araa] {10.1146/annurev-astro-082214-122414}, \href
  {https://ui.adsabs.harvard.edu/abs/2015ARA&A..53..501A} {53, 501}

\bibitem[\protect\citeauthoryear{{Andr{\'e}}, {Di Francesco}, {Ward-Thompson},
  {Inutsuka}, {Pudritz}  \& {Pineda}}{{Andr{\'e}} et~al.}{2014}]{Andre2014}
{Andr{\'e}} P.,  {Di Francesco} J.,  {Ward-Thompson} D.,  {Inutsuka} S.~I.,
  {Pudritz} R.~E.,   {Pineda} J.~E.,  2014, in {Beuther} H.,  {Klessen} R.~S.,
  {Dullemond} C.~P.,   {Henning} T.,  eds, Protostars and Planets VI. pp 27--51
  (\mn@eprint {arXiv} {1312.6232}),
  \mn@doi{10.2458/azu_uapress_9780816531240-ch002}

\bibitem[\protect\citeauthoryear{{Arce}, {Borkin}, {Goodman}, {Pineda}  \&
  {Halle}}{{Arce} et~al.}{2010}]{Arce2010}
{Arce} H.~G.,  {Borkin} M.~A.,  {Goodman} A.~A.,  {Pineda} J.~E.,   {Halle}
  M.~W.,  2010, \mn@doi [\apj] {10.1088/0004-637X/715/2/1170}, \href
  {https://ui.adsabs.harvard.edu/abs/2010ApJ...715.1170A} {715, 1170}

\bibitem[\protect\citeauthoryear{{Arzoumanian} et~al.,}{{Arzoumanian}
  et~al.}{2011}]{Arzoumanian2011}
{Arzoumanian} D.,  et~al., 2011, \mn@doi [\aap] {10.1051/0004-6361/201116596},
  \href {https://ui.adsabs.harvard.edu/abs/2011A&A...529L...6A} {529, L6}

\bibitem[\protect\citeauthoryear{{Arzoumanian} et~al.,}{{Arzoumanian}
  et~al.}{2019}]{Arzoumanian2019}
{Arzoumanian} D.,  et~al., 2019, \mn@doi [\aap] {10.1051/0004-6361/201832725},
  \href {https://ui.adsabs.harvard.edu/abs/2019A&A...621A..42A} {621, A42}

\bibitem[\protect\citeauthoryear{{Bergin} \& {Tafalla}}{{Bergin} \&
  {Tafalla}}{2007}]{BerginTafalla2007}
{Bergin} E.~A.,  {Tafalla} M.,  2007, \mn@doi [\araa]
  {10.1146/annurev.astro.45.071206.100404}, \href
  {https://ui.adsabs.harvard.edu/abs/2007ARA&A..45..339B} {45, 339}

\bibitem[\protect\citeauthoryear{Beucher \& Meyer}{Beucher \&
  Meyer}{1993}]{Beucher1993}
Beucher S.,  Meyer F.,  1993, Optical Engineering, 34, 433

\bibitem[\protect\citeauthoryear{{Bonne} et~al.,}{{Bonne}
  et~al.}{2020}]{Bonne2020}
{Bonne} L.,  et~al., 2020, \mn@doi [\aap] {10.1051/0004-6361/202038281}, \href
  {https://ui.adsabs.harvard.edu/abs/2020A&A...644A..27B} {644, A27}

\bibitem[\protect\citeauthoryear{{Briggs}}{{Briggs}}{1995}]{briggs}
{Briggs} D.~S.,  1995, PhD thesis, New Mexico Institute of Mining and
  Technology

\bibitem[\protect\citeauthoryear{{Burnham} \& {Anderson}}{{Burnham} \&
  {Anderson}}{2004}]{Burnham2004}
{Burnham} K.~P.,  {Anderson} D.~R.,  2004, Sociological Methods Research, 33,
  261

\bibitem[\protect\citeauthoryear{{CASA Team} et~al.,}{{CASA Team}
  et~al.}{2022}]{casa}
{CASA Team} et~al., 2022, \mn@doi [\pasp] {10.1088/1538-3873/ac9642}, \href
  {https://ui.adsabs.harvard.edu/abs/2022PASP..134k4501C} {134, 114501}

\bibitem[\protect\citeauthoryear{{Chapin}, {Berry}, {Gibb}, {Jenness}, {Scott},
  {Tilanus}, {Economou}  \& {Holland}}{{Chapin} et~al.}{2013}]{Chapin2013}
{Chapin} E.~L.,  {Berry} D.~S.,  {Gibb} A.~G.,  {Jenness} T.,  {Scott} D.,
  {Tilanus} R. P.~J.,  {Economou} F.,   {Holland} W.~S.,  2013, \mn@doi
  [\mnras] {10.1093/mnras/stt052}, \href
  {https://ui.adsabs.harvard.edu/abs/2013MNRAS.430.2545C} {430, 2545}

\bibitem[\protect\citeauthoryear{{Chen} \& {Ostriker}}{{Chen} \&
  {Ostriker}}{2014}]{ChenCY2014}
{Chen} C.-Y.,  {Ostriker} E.~C.,  2014, \mn@doi [\apj]
  {10.1088/0004-637X/785/1/69}, \href
  {https://ui.adsabs.harvard.edu/abs/2014ApJ...785...69C} {785, 69}

\bibitem[\protect\citeauthoryear{{Chen} \& {Ostriker}}{{Chen} \&
  {Ostriker}}{2015}]{ChenCY2015}
{Chen} C.-Y.,  {Ostriker} E.~C.,  2015, \mn@doi [\apj]
  {10.1088/0004-637X/810/2/126}, \href
  {https://ui.adsabs.harvard.edu/abs/2015ApJ...810..126C} {810, 126}

\bibitem[\protect\citeauthoryear{{Chen} \& {Ostriker}}{{Chen} \&
  {Ostriker}}{2018}]{ChenCY2018}
{Chen} C.-Y.,  {Ostriker} E.~C.,  2018, \mn@doi [\apj]
  {10.3847/1538-4357/aad905}, \href
  {https://ui.adsabs.harvard.edu/abs/2018ApJ...865...34C} {865, 34}

\bibitem[\protect\citeauthoryear{{Chen} et~al.,}{{Chen}
  et~al.}{2016a}]{ChenMike2016}
{Chen} M. C.-Y.,  et~al., 2016a, \mn@doi [\apj] {10.3847/0004-637X/826/1/95},
  \href {https://ui.adsabs.harvard.edu/abs/2016ApJ...826...95C} {826, 95}

\bibitem[\protect\citeauthoryear{{Chen}, {King}  \& {Li}}{{Chen}
  et~al.}{2016b}]{ChenCY2016}
{Chen} C.-Y.,  {King} P.~K.,   {Li} Z.-Y.,  2016b, \mn@doi [\apj]
  {10.3847/0004-637X/829/2/84}, \href
  {https://ui.adsabs.harvard.edu/abs/2016ApJ...829...84C} {829, 84}

\bibitem[\protect\citeauthoryear{{Chen}, {Mundy}, {Ostriker}, {Storm}  \&
  {Dhabal}}{{Chen} et~al.}{2020a}]{ChenCY2020}
{Chen} C.-Y.,  {Mundy} L.~G.,  {Ostriker} E.~C.,  {Storm} S.,   {Dhabal} A.,
  2020a, \mn@doi [\mnras] {10.1093/mnras/staa960}, \href
  {https://ui.adsabs.harvard.edu/abs/2020MNRAS.494.3675C} {494, 3675}

\bibitem[\protect\citeauthoryear{{Chen} et~al.,}{{Chen}
  et~al.}{2020b}]{ChenMike2020}
{Chen} M. C.-Y.,  et~al., 2020b, \mn@doi [\apj] {10.3847/1538-4357/ab7378},
  \href {https://ui.adsabs.harvard.edu/abs/2020ApJ...891...84C} {891, 84}

\bibitem[\protect\citeauthoryear{{Chen}, {Di Francesco}, {Pineda}, {Offner}  \&
  {Friesen}}{{Chen} et~al.}{2022}]{ChenMike2022}
{Chen} M. C.-Y.,  {Di Francesco} J.,  {Pineda} J.~E.,  {Offner} S. S.~R.,
  {Friesen} R.~K.,  2022, \mn@doi [\apj] {10.3847/1538-4357/ac7d4a}, \href
  {https://ui.adsabs.harvard.edu/abs/2022ApJ...935...57C} {935, 57}

\bibitem[\protect\citeauthoryear{{Choudhury} et~al.,}{{Choudhury}
  et~al.}{2020}]{Choudhury2020}
{Choudhury} S.,  et~al., 2020, \mn@doi [\aap] {10.1051/0004-6361/202037955},
  \href {https://ui.adsabs.harvard.edu/abs/2020A&A...640L...6C} {640, L6}

\bibitem[\protect\citeauthoryear{{Clarke}, {Whitworth}  \& {Hubber}}{{Clarke}
  et~al.}{2016}]{Clarke2016}
{Clarke} S.~D.,  {Whitworth} A.~P.,   {Hubber} D.~A.,  2016, \mn@doi [\mnras]
  {10.1093/mnras/stw407}, \href
  {https://ui.adsabs.harvard.edu/abs/2016MNRAS.458..319C} {458, 319}

\bibitem[\protect\citeauthoryear{{Clarke}, {Whitworth}, {Spowage},
  {Duarte-Cabral}, {Suri}, {Jaffa}, {Walch}  \& {Clark}}{{Clarke}
  et~al.}{2018}]{Clarke2018}
{Clarke} S.~D.,  {Whitworth} A.~P.,  {Spowage} R.~L.,  {Duarte-Cabral} A.,
  {Suri} S.~T.,  {Jaffa} S.~E.,  {Walch} S.,   {Clark} P.~C.,  2018, \mn@doi
  [\mnras] {10.1093/mnras/sty1675}, \href
  {https://ui.adsabs.harvard.edu/abs/2018MNRAS.479.1722C} {479, 1722}

\bibitem[\protect\citeauthoryear{{Dhabal}, {Mundy}, {Chen}, {Teuben}  \&
  {Storm}}{{Dhabal} et~al.}{2019}]{Dhabal2019}
{Dhabal} A.,  {Mundy} L.~G.,  {Chen} C.-y.,  {Teuben} P.,   {Storm} S.,  2019,
  \mn@doi [\apj] {10.3847/1538-4357/ab15d3}, \href
  {https://ui.adsabs.harvard.edu/abs/2019ApJ...876..108D} {876, 108}

\bibitem[\protect\citeauthoryear{{Doi} et~al.,}{{Doi} et~al.}{2020}]{Doi2020}
{Doi} Y.,  et~al., 2020, \mn@doi [\apj] {10.3847/1538-4357/aba1e2}, \href
  {https://ui.adsabs.harvard.edu/abs/2020ApJ...899...28D} {899, 28}

\bibitem[\protect\citeauthoryear{{Doi} et~al.,}{{Doi} et~al.}{2021}]{Doi2021}
{Doi} Y.,  et~al., 2021, \mn@doi [\apjl] {10.3847/2041-8213/ac3cc1}, \href
  {https://ui.adsabs.harvard.edu/abs/2021ApJ...923L...9D} {923, L9}

\bibitem[\protect\citeauthoryear{{Drabek} et~al.,}{{Drabek}
  et~al.}{2012}]{Drabek2012}
{Drabek} E.,  et~al., 2012, \mn@doi [\mnras]
  {10.1111/j.1365-2966.2012.21140.x}, \href
  {https://ui.adsabs.harvard.edu/abs/2012MNRAS.426...23D} {426, 23}

\bibitem[\protect\citeauthoryear{{Dunham} et~al.,}{{Dunham}
  et~al.}{2015}]{Dunham2015}
{Dunham} M.~M.,  et~al., 2015, \mn@doi [\apjs] {10.1088/0067-0049/220/1/11},
  \href {https://ui.adsabs.harvard.edu/abs/2015ApJS..220...11D} {220, 11}

\bibitem[\protect\citeauthoryear{{Durand} \& {Greenwood}}{{Durand} \&
  {Greenwood}}{1958}]{Durand1958}
{Durand} D.,  {Greenwood} J.~A.,  1958, \mn@doi [Journal of Geology]
  {10.1086/626501}, \href
  {https://ui.adsabs.harvard.edu/abs/1958JG.....66..229D} {66, 229}

\bibitem[\protect\citeauthoryear{Ester, Kriegel, Sander  \& Xu}{Ester
  et~al.}{1996}]{Ester1996}
Ester M.,  Kriegel H.-P.,  Sander J.,   Xu X.,  1996, in Proc. of 2nd
  International Conference on Knowledge Discovery and. pp 226--231

\bibitem[\protect\citeauthoryear{{Federrath}}{{Federrath}}{2016}]{Federrath2016}
{Federrath} C.,  2016, \mn@doi [\mnras] {10.1093/mnras/stv2880}, \href
  {https://ui.adsabs.harvard.edu/abs/2016MNRAS.457..375F} {457, 375}

\bibitem[\protect\citeauthoryear{{Fissel} et~al.,}{{Fissel}
  et~al.}{2019}]{Fissel2019}
{Fissel} L.~M.,  et~al., 2019, \mn@doi [\apj] {10.3847/1538-4357/ab1eb0}, \href
  {https://ui.adsabs.harvard.edu/abs/2019ApJ...878..110F} {878, 110}

\bibitem[\protect\citeauthoryear{{Friberg}, {Bastien}, {Berry}, {Savini},
  {Graves}  \& {Pattle}}{{Friberg} et~al.}{2016}]{Friberg2016}
{Friberg} P.,  {Bastien} P.,  {Berry} D.,  {Savini} G.,  {Graves} S.~F.,
  {Pattle} K.,  2016, in {Holland} W.~S.,  {Zmuidzinas} J.,  eds,  Society of
  Photo-Optical Instrumentation Engineers (SPIE) Conference Series Vol. 9914,
  Millimeter, Submillimeter, and Far-Infrared Detectors and Instrumentation for
  Astronomy VIII. p. 991403, \mn@doi{10.1117/12.2231943}

\bibitem[\protect\citeauthoryear{{Friesen}, {Medeiros}, {Schnee}, {Bourke}, {di
  Francesco}, {Gutermuth}  \& {Myers}}{{Friesen} et~al.}{2013}]{Friesen2013}
{Friesen} R.~K.,  {Medeiros} L.,  {Schnee} S.,  {Bourke} T.~L.,  {di Francesco}
  J.,  {Gutermuth} R.,   {Myers} P.~C.,  2013, \mn@doi [\mnras]
  {10.1093/mnras/stt1671}, \href
  {https://ui.adsabs.harvard.edu/abs/2013MNRAS.436.1513F} {436, 1513}

\bibitem[\protect\citeauthoryear{{Friesen} et~al.,}{{Friesen}
  et~al.}{2017}]{GAS}
{Friesen} R.~K.,  et~al., 2017, \mn@doi [\apj] {10.3847/1538-4357/aa6d58},
  \href {https://ui.adsabs.harvard.edu/abs/2017ApJ...843...63F} {843, 63}

\bibitem[\protect\citeauthoryear{{Ginsburg}, {Sokolov}, {de Val-Borro},
  {Rosolowsky}, {Pineda}, {Sip{\H{o}}cz}  \& {Henshaw}}{{Ginsburg}
  et~al.}{2022}]{Ginsburg2022}
{Ginsburg} A.,  {Sokolov} V.,  {de Val-Borro} M.,  {Rosolowsky} E.,  {Pineda}
  J.~E.,  {Sip{\H{o}}cz} B.~M.,   {Henshaw} J.~D.,  2022, \mn@doi [\aj]
  {10.3847/1538-3881/ac695a}, \href
  {https://ui.adsabs.harvard.edu/abs/2022AJ....163..291G} {163, 291}

\bibitem[\protect\citeauthoryear{{G{\'o}mez}, {V{\'a}zquez-Semadeni}  \&
  {Zamora-Avil{\'e}s}}{{G{\'o}mez} et~al.}{2018}]{Gomez2018}
{G{\'o}mez} G.~C.,  {V{\'a}zquez-Semadeni} E.,   {Zamora-Avil{\'e}s} M.,  2018,
  \mn@doi [\mnras] {10.1093/mnras/sty2018}, \href
  {https://ui.adsabs.harvard.edu/abs/2018MNRAS.480.2939G} {480, 2939}

\bibitem[\protect\citeauthoryear{{Hacar}, {Tafalla}, {Kauffmann}  \&
  {Kov{\'a}cs}}{{Hacar} et~al.}{2013}]{Hacar2013}
{Hacar} A.,  {Tafalla} M.,  {Kauffmann} J.,   {Kov{\'a}cs} A.,  2013, \mn@doi
  [\aap] {10.1051/0004-6361/201220090}, \href
  {https://ui.adsabs.harvard.edu/abs/2013A&A...554A..55H} {554, A55}

\bibitem[\protect\citeauthoryear{{Hacar}, {Tafalla}  \& {Alves}}{{Hacar}
  et~al.}{2017}]{Hacar2017}
{Hacar} A.,  {Tafalla} M.,   {Alves} J.,  2017, \mn@doi [\aap]
  {10.1051/0004-6361/201630348}, \href
  {https://ui.adsabs.harvard.edu/abs/2017A&A...606A.123H} {606, A123}

\bibitem[\protect\citeauthoryear{{Hacar}, {Clark}, {Heitsch}, {Kainulainen},
  {Panopoulou}, {Seifried}  \& {Smith}}{{Hacar} et~al.}{2023}]{Hacar2023}
{Hacar} A.,  {Clark} S.~E.,  {Heitsch} F.,  {Kainulainen} J.,  {Panopoulou}
  G.~V.,  {Seifried} D.,   {Smith} R.,  2023, in {Inutsuka} S.,  {Aikawa} Y.,
  {Muto} T.,  {Tomida} K.,   {Tamura} M.,  eds,  Astronomical Society of the
  Pacific Conference Series Vol. 534, Protostars and Planets VII. p.~153
  (\mn@eprint {arXiv} {2203.09562}), \mn@doi{10.48550/arXiv.2203.09562}

\bibitem[\protect\citeauthoryear{{Harvey}, {Wilking}  \& {Joy}}{{Harvey}
  et~al.}{1984}]{Harvey1984}
{Harvey} P.~M.,  {Wilking} B.~A.,   {Joy} M.,  1984, \mn@doi [\apj]
  {10.1086/161777}, \href
  {https://ui.adsabs.harvard.edu/abs/1984ApJ...278..156H} {278, 156}

\bibitem[\protect\citeauthoryear{{Heigl}, {Gritschneder}  \& {Burkert}}{{Heigl}
  et~al.}{2020}]{Heigl2020}
{Heigl} S.,  {Gritschneder} M.,   {Burkert} A.,  2020, \mn@doi [\mnras]
  {10.1093/mnras/staa1202}, \href
  {https://ui.adsabs.harvard.edu/abs/2020MNRAS.495..758H} {495, 758}

\bibitem[\protect\citeauthoryear{{Heyvaerts} \& {Priest}}{{Heyvaerts} \&
  {Priest}}{1983}]{Heyvaerts1983}
{Heyvaerts} J.,  {Priest} E.~R.,  1983, \aap, \href
  {https://ui.adsabs.harvard.edu/abs/1983A&A...117..220H} {117, 220}

\bibitem[\protect\citeauthoryear{{Inoue}, {Hennebelle}, {Fukui}, {Matsumoto},
  {Iwasaki}  \& {Inutsuka}}{{Inoue} et~al.}{2018}]{Inoue2018}
{Inoue} T.,  {Hennebelle} P.,  {Fukui} Y.,  {Matsumoto} T.,  {Iwasaki} K.,
  {Inutsuka} S.-i.,  2018, \mn@doi [\pasj] {10.1093/pasj/psx089}, \href
  {https://ui.adsabs.harvard.edu/abs/2018PASJ...70S..53I} {70, S53}

\bibitem[\protect\citeauthoryear{{Inutsuka}, {Inoue}, {Iwasaki}  \&
  {Hosokawa}}{{Inutsuka} et~al.}{2015}]{Inutsuka2015}
{Inutsuka} S.-i.,  {Inoue} T.,  {Iwasaki} K.,   {Hosokawa} T.,  2015, \mn@doi
  [\aap] {10.1051/0004-6361/201425584}, \href
  {https://ui.adsabs.harvard.edu/abs/2015A&A...580A..49I} {580, A49}

\bibitem[\protect\citeauthoryear{{Jow}, {Hill}, {Scott}, {Soler}, {Martin},
  {Devlin}, {Fissel}  \& {Poidevin}}{{Jow} et~al.}{2018}]{Jow2018}
{Jow} D.~L.,  {Hill} R.,  {Scott} D.,  {Soler} J.~D.,  {Martin} P.~G.,
  {Devlin} M.~J.,  {Fissel} L.~M.,   {Poidevin} F.,  2018, \mn@doi [\mnras]
  {10.1093/mnras/stx2736}, \href
  {https://ui.adsabs.harvard.edu/abs/2018MNRAS.474.1018J} {474, 1018}

\bibitem[\protect\citeauthoryear{{Kirk}, {Johnstone}  \& {Tafalla}}{{Kirk}
  et~al.}{2007}]{KirkHelen2007}
{Kirk} H.,  {Johnstone} D.,   {Tafalla} M.,  2007, \mn@doi [\apj]
  {10.1086/521395}, \href
  {https://ui.adsabs.harvard.edu/abs/2007ApJ...668.1042K} {668, 1042}

\bibitem[\protect\citeauthoryear{{Kirk}, {Myers}, {Bourke}, {Gutermuth},
  {Hedden}  \& {Wilson}}{{Kirk} et~al.}{2013}]{Kirk2013}
{Kirk} H.,  {Myers} P.~C.,  {Bourke} T.~L.,  {Gutermuth} R.~A.,  {Hedden} A.,
  {Wilson} G.~W.,  2013, \mn@doi [\apj] {10.1088/0004-637X/766/2/115}, \href
  {https://ui.adsabs.harvard.edu/abs/2013ApJ...766..115K} {766, 115}

\bibitem[\protect\citeauthoryear{{Kirk} et~al.,}{{Kirk}
  et~al.}{2018}]{KirkH2018}
{Kirk} H.,  et~al., 2018, \mn@doi [\apjs] {10.3847/1538-4365/aada7f}, \href
  {https://ui.adsabs.harvard.edu/abs/2018ApJS..238....8K} {238, 8}

\bibitem[\protect\citeauthoryear{{Klessen} \& {Hennebelle}}{{Klessen} \&
  {Hennebelle}}{2010}]{Klessen2010}
{Klessen} R.~S.,  {Hennebelle} P.,  2010, \mn@doi [\aap]
  {10.1051/0004-6361/200913780}, \href
  {https://ui.adsabs.harvard.edu/abs/2010A&A...520A..17K} {520, A17}

\bibitem[\protect\citeauthoryear{{Koch} \& {Ginsburg}}{{Koch} \&
  {Ginsburg}}{2022}]{uvcombine}
{Koch} E.,  {Ginsburg} A.,  2022, {uvcombine: Combine images with different
  resolutions}, Astrophysics Source Code Library, record ascl:2208.014
  (\mn@eprint {ascl} {2208.014})

\bibitem[\protect\citeauthoryear{{Lee} et~al.,}{{Lee} et~al.}{2021}]{LeeD2021}
{Lee} D.,  et~al., 2021, \mn@doi [\apj] {10.3847/1538-4357/ac0cf2}, \href
  {https://ui.adsabs.harvard.edu/abs/2021ApJ...918...39L} {918, 39}

\bibitem[\protect\citeauthoryear{{Levenberg}}{{Levenberg}}{1944}]{Levenberg1944}
{Levenberg} K.,  1944, Quarterly of Applied Mathematics, 2, 164

\bibitem[\protect\citeauthoryear{Marquardt}{Marquardt}{1963}]{Marquardt1963}
Marquardt D.~W.,  1963, \mn@doi [SIAM Journal on Applied Mathematics]
  {10.1137/0111030}, 11, 431

\bibitem[\protect\citeauthoryear{{Monsch} et~al.,}{{Monsch}
  et~al.}{2018}]{Monsch2018}
{Monsch} K.,  et~al., 2018, \mn@doi [\apj] {10.3847/1538-4357/aac8da}, \href
  {https://ui.adsabs.harvard.edu/abs/2018ApJ...861...77M} {861, 77}

\bibitem[\protect\citeauthoryear{{Mor{\'e}}}{{Mor{\'e}}}{1978}]{More1978}
{Mor{\'e}} J.~J.,  1978, \mn@doi [Lecture Notes in Mathematics, Berlin Springer
  Verlag] {10.1007/BFb0067700}, \href
  {http://adsabs.harvard.edu/abs/1978LNM...630..105M} {630, 105}

\bibitem[\protect\citeauthoryear{{Nagai}, {Inutsuka}  \& {Miyama}}{{Nagai}
  et~al.}{1998}]{Nagai1998}
{Nagai} T.,  {Inutsuka} S.-i.,   {Miyama} S.~M.,  1998, \mn@doi [\apj]
  {10.1086/306249}, \href
  {https://ui.adsabs.harvard.edu/abs/1998ApJ...506..306N} {506, 306}

\bibitem[\protect\citeauthoryear{{Naghizadeh-Khouei} \&
  {Clarke}}{{Naghizadeh-Khouei} \& {Clarke}}{1993}]{Naghizadeh-Khouei1993}
{Naghizadeh-Khouei} J.,  {Clarke} D.,  1993, \aap, \href
  {https://ui.adsabs.harvard.edu/abs/1993A&A...274..968N} {274, 968}

\bibitem[\protect\citeauthoryear{{Nakamura} \& {Li}}{{Nakamura} \&
  {Li}}{2008}]{Nakamura2008}
{Nakamura} F.,  {Li} Z.-Y.,  2008, \mn@doi [\apj] {10.1086/591641}, \href
  {https://ui.adsabs.harvard.edu/abs/2008ApJ...687..354N} {687, 354}

\bibitem[\protect\citeauthoryear{{Nakano} \& {Nakamura}}{{Nakano} \&
  {Nakamura}}{1978}]{Nakano1978}
{Nakano} T.,  {Nakamura} T.,  1978, \pasj, \href
  {https://ui.adsabs.harvard.edu/abs/1978PASJ...30..671N} {30, 671}

\bibitem[\protect\citeauthoryear{{Offner} \& {Liu}}{{Offner} \&
  {Liu}}{2018}]{Offner2018}
{Offner} S. S.~R.,  {Liu} Y.,  2018, \mn@doi [Nature Astronomy]
  {10.1038/s41550-018-0566-1}, \href
  {https://ui.adsabs.harvard.edu/abs/2018NatAs...2..896O} {2, 896}

\bibitem[\protect\citeauthoryear{{Palmeirim} et~al.,}{{Palmeirim}
  et~al.}{2013}]{Palmeirim2013}
{Palmeirim} P.,  et~al., 2013, \mn@doi [\aap] {10.1051/0004-6361/201220500},
  \href {https://ui.adsabs.harvard.edu/abs/2013A&A...550A..38P} {550, A38}

\bibitem[\protect\citeauthoryear{{Pandhi} et~al.,}{{Pandhi}
  et~al.}{2023}]{Pandhi2023}
{Pandhi} A.,  et~al., 2023, \mn@doi [\mnras] {10.1093/mnras/stad2283}, \href
  {https://ui.adsabs.harvard.edu/abs/2023MNRAS.525..364P} {525, 364}

\bibitem[\protect\citeauthoryear{{Parsons} et~al.,}{{Parsons}
  et~al.}{2018}]{Parsons2018}
{Parsons} H.,  et~al., 2018, \mn@doi [\apjs] {10.3847/1538-4365/aa989c}, \href
  {https://ui.adsabs.harvard.edu/abs/2018ApJS..234...22P} {234, 22}

\bibitem[\protect\citeauthoryear{{Pattle}, {Fissel}, {Tahani}, {Liu}  \&
  {Ntormousi}}{{Pattle} et~al.}{2023}]{Pattle2023}
{Pattle} K.,  {Fissel} L.,  {Tahani} M.,  {Liu} T.,   {Ntormousi} E.,  2023, in
  {Inutsuka} S.,  {Aikawa} Y.,  {Muto} T.,  {Tomida} K.,   {Tamura} M.,  eds,
  Astronomical Society of the Pacific Conference Series Vol. 534, Protostars
  and Planets VII. p.~193 (\mn@eprint {arXiv} {2203.11179}),
  \mn@doi{10.48550/arXiv.2203.11179}

\bibitem[\protect\citeauthoryear{Pedregosa et~al.,}{Pedregosa
  et~al.}{2011}]{scikit-learn}
Pedregosa F.,  et~al., 2011, Journal of Machine Learning Research, 12, 2825

\bibitem[\protect\citeauthoryear{{Peretto}, {Andr{\'e}}  \&
  {Belloche}}{{Peretto} et~al.}{2006}]{Peretto2006}
{Peretto} N.,  {Andr{\'e}} P.,   {Belloche} A.,  2006, \mn@doi [\aap]
  {10.1051/0004-6361:20053324}, \href
  {https://ui.adsabs.harvard.edu/abs/2006A&A...445..979P} {445, 979}

\bibitem[\protect\citeauthoryear{{Pezzuto} et~al.,}{{Pezzuto}
  et~al.}{2021}]{Pezzuto2021}
{Pezzuto} S.,  et~al., 2021, \mn@doi [\aap] {10.1051/0004-6361/201936534},
  \href {https://ui.adsabs.harvard.edu/abs/2021A&A...645A..55P} {645, A55}

\bibitem[\protect\citeauthoryear{{Pillai} et~al.,}{{Pillai}
  et~al.}{2020}]{Pillai2020}
{Pillai} T. G.~S.,  et~al., 2020, \mn@doi [Nature Astronomy]
  {10.1038/s41550-020-1172-6}, \href
  {https://ui.adsabs.harvard.edu/abs/2020NatAs...4.1195P} {4, 1195}

\bibitem[\protect\citeauthoryear{{Pineda}, {Goodman}, {Arce}, {Caselli},
  {Foster}, {Myers}  \& {Rosolowsky}}{{Pineda} et~al.}{2010}]{Pineda2010}
{Pineda} J.~E.,  {Goodman} A.~A.,  {Arce} H.~G.,  {Caselli} P.,  {Foster}
  J.~B.,  {Myers} P.~C.,   {Rosolowsky} E.~W.,  2010, \mn@doi [\apjl]
  {10.1088/2041-8205/712/1/L116}, \href
  {https://ui.adsabs.harvard.edu/abs/2010ApJ...712L.116P} {712, L116}

\bibitem[\protect\citeauthoryear{{Pineda} et~al.,}{{Pineda}
  et~al.}{2023}]{Pineda2023}
{Pineda} J.~E.,  et~al., 2023, in {Inutsuka} S.,  {Aikawa} Y.,  {Muto} T.,
  {Tomida} K.,   {Tamura} M.,  eds,  Astronomical Society of the Pacific
  Conference Series Vol. 534, Protostars and Planets VII. p.~233 (\mn@eprint
  {arXiv} {2205.03935}), \mn@doi{10.48550/arXiv.2205.03935}

\bibitem[\protect\citeauthoryear{{Planck Collaboration} et~al.,}{{Planck
  Collaboration} et~al.}{2016}]{Planck2016_HROMagClouds}
{Planck Collaboration} et~al., 2016, \mn@doi [\aap]
  {10.1051/0004-6361/201525896}, \href
  {https://ui.adsabs.harvard.edu/abs/2016A&A...586A.138P} {586, A138}

\bibitem[\protect\citeauthoryear{{Plunkett} et~al.,}{{Plunkett}
  et~al.}{2023}]{SDcomb}
{Plunkett} A.,  et~al., 2023, \mn@doi [\pasp] {10.1088/1538-3873/acb9bd}, \href
  {https://ui.adsabs.harvard.edu/abs/2023PASP..135c4501P} {135, 034501}

\bibitem[\protect\citeauthoryear{{Pon}, {Johnstone}  \& {Heitsch}}{{Pon}
  et~al.}{2011}]{Pon2011}
{Pon} A.,  {Johnstone} D.,   {Heitsch} F.,  2011, \mn@doi [\apj]
  {10.1088/0004-637X/740/2/88}, \href
  {https://ui.adsabs.harvard.edu/abs/2011ApJ...740...88P} {740, 88}

\bibitem[\protect\citeauthoryear{{Porter}, {Pouquet}  \& {Woodward}}{{Porter}
  et~al.}{1994}]{Porter1994}
{Porter} D.~H.,  {Pouquet} A.,   {Woodward} P.~R.,  1994, \mn@doi [Physics of
  Fluids] {10.1063/1.868217}, \href
  {https://ui.adsabs.harvard.edu/abs/1994PhFl....6.2133P} {6, 2133}

\bibitem[\protect\citeauthoryear{{Seifried} \& {Walch}}{{Seifried} \&
  {Walch}}{2015}]{Seifried2015}
{Seifried} D.,  {Walch} S.,  2015, \mn@doi [\mnras] {10.1093/mnras/stv1458},
  \href {https://ui.adsabs.harvard.edu/abs/2015MNRAS.452.2410S} {452, 2410}

\bibitem[\protect\citeauthoryear{{Shimajiri}, {Andr{\'e}}, {Palmeirim},
  {Arzoumanian}, {Bracco}, {K{\"o}nyves}, {Ntormousi}  \&
  {Ladjelate}}{{Shimajiri} et~al.}{2019}]{Shimajiri2019}
{Shimajiri} Y.,  {Andr{\'e}} P.,  {Palmeirim} P.,  {Arzoumanian} D.,  {Bracco}
  A.,  {K{\"o}nyves} V.,  {Ntormousi} E.,   {Ladjelate} B.,  2019, \mn@doi
  [\aap] {10.1051/0004-6361/201834399}, \href
  {https://ui.adsabs.harvard.edu/abs/2019A&A...623A..16S} {623, A16}

\bibitem[\protect\citeauthoryear{{Shirley}}{{Shirley}}{2015}]{Shirley2015}
{Shirley} Y.~L.,  2015, \mn@doi [\pasp] {10.1086/680342}, \href
  {https://ui.adsabs.harvard.edu/abs/2015PASP..127..299S} {127, 299}

\bibitem[\protect\citeauthoryear{{Smith}, {Glover}, {Klessen}  \&
  {Fuller}}{{Smith} et~al.}{2016}]{SmithR2016}
{Smith} R.~J.,  {Glover} S. C.~O.,  {Klessen} R.~S.,   {Fuller} G.~A.,  2016,
  \mn@doi [\mnras] {10.1093/mnras/stv2559}, \href
  {https://ui.adsabs.harvard.edu/abs/2016MNRAS.455.3640S} {455, 3640}

\bibitem[\protect\citeauthoryear{{Soam} et~al.,}{{Soam}
  et~al.}{2019}]{Soam2019}
{Soam} A.,  et~al., 2019, \mn@doi [\apj] {10.3847/1538-4357/ab39dd}, \href
  {https://ui.adsabs.harvard.edu/abs/2019ApJ...883...95S} {883, 95}

\bibitem[\protect\citeauthoryear{{Sokolov}, {Pineda}, {Buchner}  \&
  {Caselli}}{{Sokolov} et~al.}{2020}]{Sokolov2020}
{Sokolov} V.,  {Pineda} J.~E.,  {Buchner} J.,   {Caselli} P.,  2020, \mn@doi
  [\apjl] {10.3847/2041-8213/ab8018}, \href
  {https://ui.adsabs.harvard.edu/abs/2020ApJ...892L..32S} {892, L32}

\bibitem[\protect\citeauthoryear{{Soler}}{{Soler}}{2019}]{Soler2019}
{Soler} J.~D.,  2019, \mn@doi [\aap] {10.1051/0004-6361/201935779}, \href
  {https://ui.adsabs.harvard.edu/abs/2019A&A...629A..96S} {629, A96}

\bibitem[\protect\citeauthoryear{{Soler} \& {Hennebelle}}{{Soler} \&
  {Hennebelle}}{2017}]{Soler2017}
{Soler} J.~D.,  {Hennebelle} P.,  2017, \mn@doi [\aap]
  {10.1051/0004-6361/201731049}, \href
  {https://ui.adsabs.harvard.edu/abs/2017A&A...607A...2S} {607, A2}

\bibitem[\protect\citeauthoryear{{Soler}, {Hennebelle}, {Martin},
  {Miville-Desch{\^e}nes}, {Netterfield}  \& {Fissel}}{{Soler}
  et~al.}{2013}]{Soler2013}
{Soler} J.~D.,  {Hennebelle} P.,  {Martin} P.~G.,  {Miville-Desch{\^e}nes}
  M.~A.,  {Netterfield} C.~B.,   {Fissel} L.~M.,  2013, \mn@doi [\apj]
  {10.1088/0004-637X/774/2/128}, \href
  {https://ui.adsabs.harvard.edu/abs/2013ApJ...774..128S} {774, 128}

\bibitem[\protect\citeauthoryear{{Sugiura}}{{Sugiura}}{1978}]{Sugiura1978}
{Sugiura} N.,  1978, \mn@doi [Communications in Statistics - Theory and
  Methods] {10.1080/03610927808827599}, 7, 13

\bibitem[\protect\citeauthoryear{{Tafalla} \& {Hacar}}{{Tafalla} \&
  {Hacar}}{2015}]{Tafalla2015}
{Tafalla} M.,  {Hacar} A.,  2015, \mn@doi [\aap] {10.1051/0004-6361/201424576},
  \href {https://ui.adsabs.harvard.edu/abs/2015A&A...574A.104T} {574, A104}

\bibitem[\protect\citeauthoryear{{Tang}, {Koch}, {Peretto}, {Novak},
  {Duarte-Cabral}, {Chapman}, {Hsieh}  \& {Yen}}{{Tang}
  et~al.}{2019}]{TangYW2019}
{Tang} Y.-W.,  {Koch} P.~M.,  {Peretto} N.,  {Novak} G.,  {Duarte-Cabral} A.,
  {Chapman} N.~L.,  {Hsieh} P.-Y.,   {Yen} H.-W.,  2019, \mn@doi [\apj]
  {10.3847/1538-4357/ab1484}, \href
  {https://ui.adsabs.harvard.edu/abs/2019ApJ...878...10T} {878, 10}

\bibitem[\protect\citeauthoryear{{Tritsis} \& {Tassis}}{{Tritsis} \&
  {Tassis}}{2016}]{Tritsis2016}
{Tritsis} A.,  {Tassis} K.,  2016, \mn@doi [\mnras] {10.1093/mnras/stw1881},
  \href {https://ui.adsabs.harvard.edu/abs/2016MNRAS.462.3602T} {462, 3602}

\bibitem[\protect\citeauthoryear{{Tritsis} \& {Tassis}}{{Tritsis} \&
  {Tassis}}{2018}]{Tritsis2018}
{Tritsis} A.,  {Tassis} K.,  2018, \mn@doi [Science] {10.1126/science.aao1185},
  \href {https://ui.adsabs.harvard.edu/abs/2018Sci...360..635T} {360, 635}

\bibitem[\protect\citeauthoryear{{Vaillancourt}}{{Vaillancourt}}{2006}]{Vaillancourt2006}
{Vaillancourt} J.~E.,  2006, \mn@doi [\pasp] {10.1086/507472}, \href
  {https://ui.adsabs.harvard.edu/abs/2006PASP..118.1340V} {118, 1340}

\bibitem[\protect\citeauthoryear{{Vazquez-Semadeni}}{{Vazquez-Semadeni}}{1994}]{Vazquez-Semadeni1994}
{Vazquez-Semadeni} E.,  1994, \mn@doi [\apj] {10.1086/173847}, \href
  {https://ui.adsabs.harvard.edu/abs/1994ApJ...423..681V} {423, 681}

\bibitem[\protect\citeauthoryear{{V{\'a}zquez-Semadeni}, {Palau},
  {Ballesteros-Paredes}, {G{\'o}mez}  \&
  {Zamora-Avil{\'e}s}}{{V{\'a}zquez-Semadeni}
  et~al.}{2019}]{Vazquez-Semadeni2019}
{V{\'a}zquez-Semadeni} E.,  {Palau} A.,  {Ballesteros-Paredes} J.,  {G{\'o}mez}
  G.~C.,   {Zamora-Avil{\'e}s} M.,  2019, \mn@doi [\mnras]
  {10.1093/mnras/stz2736}, \href
  {https://ui.adsabs.harvard.edu/abs/2019MNRAS.490.3061V} {490, 3061}

\bibitem[\protect\citeauthoryear{{Walawender}, {Bally}, {Francesco},
  {J{\o}rgensen}  \& {Getman}}{{Walawender} et~al.}{2008}]{Walawender2008}
{Walawender} J.,  {Bally} J.,  {Francesco} J.~D.,  {J{\o}rgensen} J.,
  {Getman} K.~.,  2008, in {Reipurth} B.,  ed., , Vol.~4, Handbook of Star
  Forming Regions, Volume I.
p.~346

\bibitem[\protect\citeauthoryear{{Walsh}, {Bourke}  \& {Myers}}{{Walsh}
  et~al.}{2006}]{Walsh2006}
{Walsh} A.~J.,  {Bourke} T.~L.,   {Myers} P.~C.,  2006, \mn@doi [\apj]
  {10.1086/498564}, \href
  {https://ui.adsabs.harvard.edu/abs/2006ApJ...637..860W} {637, 860}

\bibitem[\protect\citeauthoryear{{Wang}, {Koch}, {Galv{\'a}n-Madrid}, {Lai},
  {Liu}, {Lin}  \& {Pattle}}{{Wang} et~al.}{2020}]{WangJW2020}
{Wang} J.-W.,  {Koch} P.~M.,  {Galv{\'a}n-Madrid} R.,  {Lai} S.-P.,  {Liu}
  H.~B.,  {Lin} S.-J.,   {Pattle} K.,  2020, \mn@doi [\apj]
  {10.3847/1538-4357/abc74e}, \href
  {https://ui.adsabs.harvard.edu/abs/2020ApJ...905..158W} {905, 158}

\bibitem[\protect\citeauthoryear{{Ward-Thompson} et~al.,}{{Ward-Thompson}
  et~al.}{2017}]{Ward-Thompson2017}
{Ward-Thompson} D.,  et~al., 2017, \mn@doi [\apj] {10.3847/1538-4357/aa70a0},
  \href {https://ui.adsabs.harvard.edu/abs/2017ApJ...842...66W} {842, 66}

\bibitem[\protect\citeauthoryear{{Wardle} \& {Kronberg}}{{Wardle} \&
  {Kronberg}}{1974}]{Wardle1974}
{Wardle} J.~F.~C.,  {Kronberg} P.~P.,  1974, \mn@doi [\apj] {10.1086/153240},
  \href {https://ui.adsabs.harvard.edu/abs/1974ApJ...194..249W} {194, 249}

\bibitem[\protect\citeauthoryear{{Zucker}, {Schlafly}, {Speagle}, {Green},
  {Portillo}, {Finkbeiner}  \& {Goodman}}{{Zucker} et~al.}{2018}]{Zucker2018}
{Zucker} C.,  {Schlafly} E.~F.,  {Speagle} J.~S.,  {Green} G.~M.,  {Portillo}
  S. K.~N.,  {Finkbeiner} D.~P.,   {Goodman} A.~A.,  2018, \mn@doi [\apj]
  {10.3847/1538-4357/aae97c}, \href
  {https://ui.adsabs.harvard.edu/abs/2018ApJ...869...83Z} {869, 83}

\makeatother
\end{thebibliography}




\appendix


\section{CO contamination}
\label{appendix:CO}

\begin{figure}
    \includegraphics[width=\columnwidth]{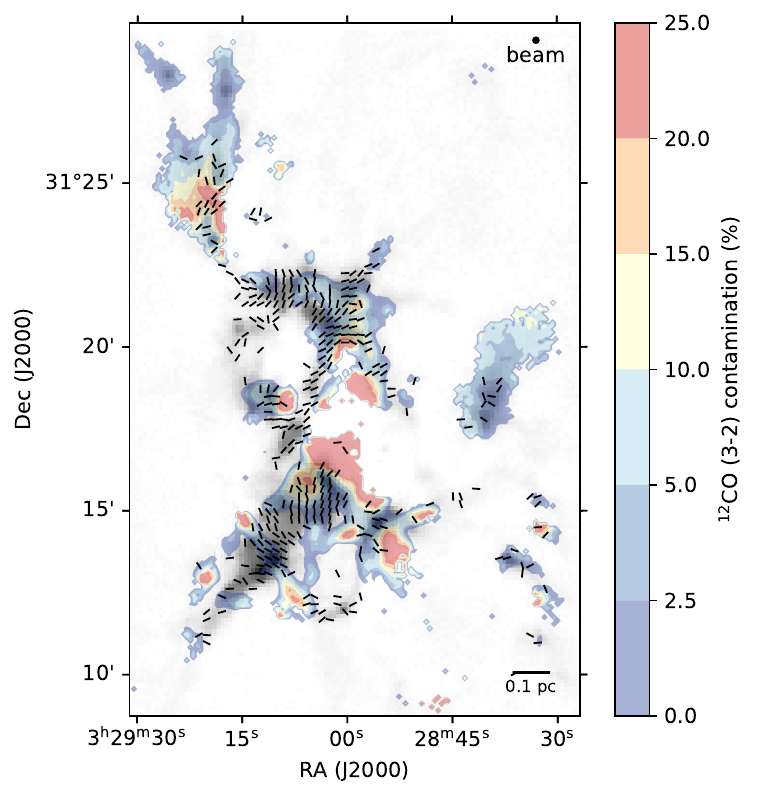}
    \caption{The CO (3 -- 2) contamination in NGC 1333 plotted over the GBS, CO-decontaminated map of NGC 1333 in greyscale.  The 850\,$\mu$m continuum contamination fraction is shown as filled contours (coloured) and is measured as a percentage of the 850 \textmu m continuum.  The inferred $B$ field is overlaid as black line segments.}
    \label{fig:CO_cont}
\end{figure}

Figure \ref{fig:CO_cont} shows the CO (3 -- 2) contamination levels in JCMT GBS's 850 \textmu m data. The contamination level is measured in integrated CO brightness as a fraction of the decontaminated 850 \textmu m emission \citep{Drabek2012}. The integrated CO emission map was filtered similarly through the same data reduction process as the 850 \textmu m map before its removal to ensure the same spatial sensitivity. CO contamination is most significant in areas where faint dust emission overlaps with bright and broad CO outflows, where the integrated CO emission is significant relative to the dust continuum.  Nevertheless, as seen in Figure \ref{fig:CO_cont}, most of our magnetic field measurements are detected over regions of bright dust emission, where the CO contamination level is low ($< 15\%$). The only exception is towards the northwestern corner of NGC 1333, where CO may have been particularly excited by the feedback of the B stars nearby.

\section{DBSCAN Results}
\label{appendix:DBScan}

\begin{figure*}
    \includegraphics[width=0.87\textwidth]{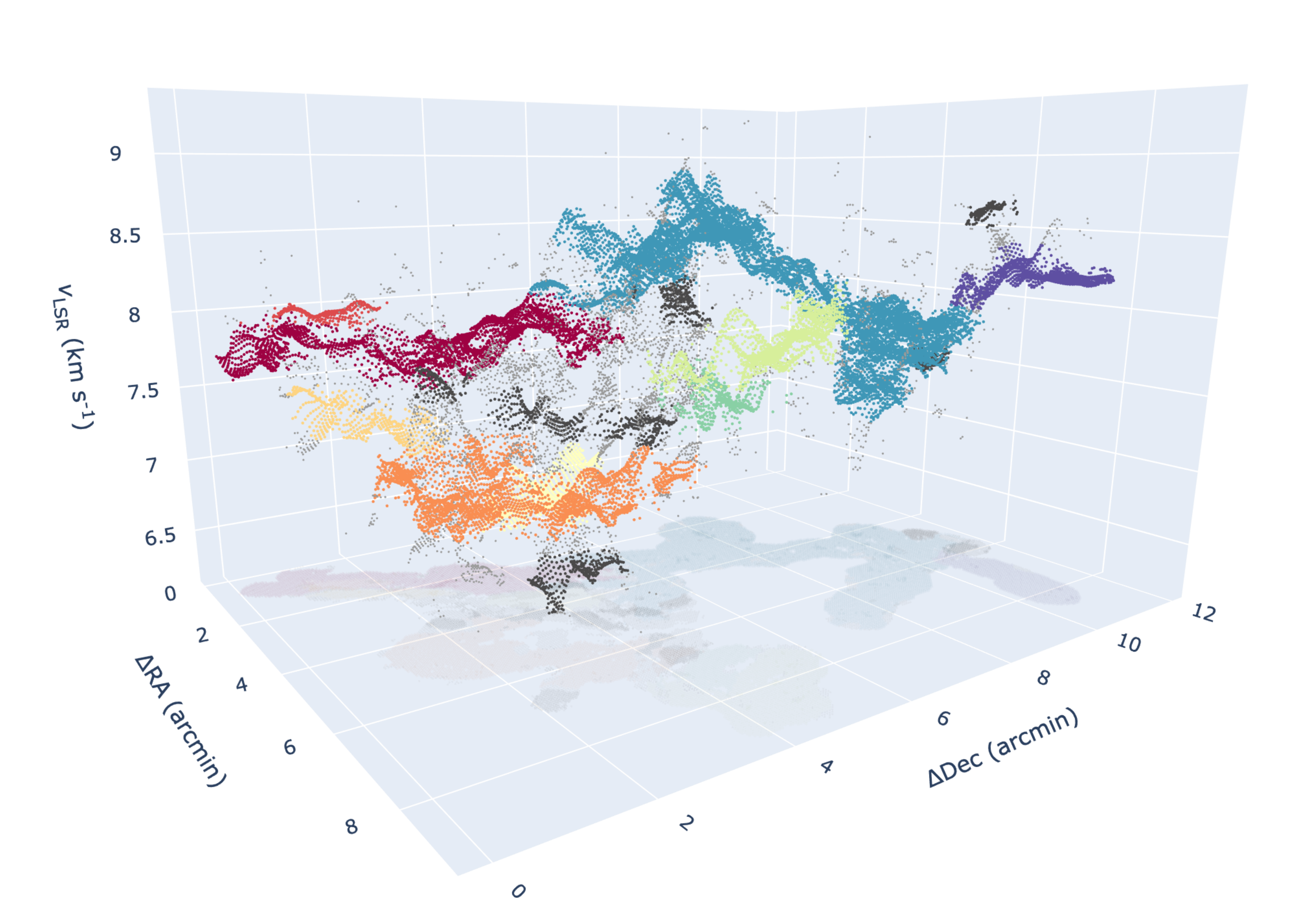}
    \caption{Same as Figure \ref{fig:dbscan_3d}, but with noise (light grey) and structures with less than 250 members (dark grey) identified by our DBSCAN run also shown.}
    \label{fig:dbscan_3d_noise}
\end{figure*}

Figure \ref{fig:dbscan_3d_noise} shows a 3D scatter plot of our full \texttt{DBSCAN} results in PPV space. The nine VCSs included in our final analyses are coloured the same as Figure \ref{fig:dbscan_3d}. Additionally, the noise-classified members and compact VCSs (with $< 250$ members) excluded from our analyses are shown in light and dark grey, respectively. As can be seen from the shadow of the structures projected onto the RA-Dec-plane, the VCSs with less than 250 members are indeed spatially compact. The noise samples, on the other hand, appear as either isolated pixels or as relatively correlated structures between the larger VCSs in PPV space. The exclusion of both the noise and the compact VCSs, however, does not affect our analysis, given that they represent structures that are distinct in the PPV space from those included in our analysis.

\section{Column Density Gradients}
\label{appendix:ColDen}

\subsection{Comparing with Dust Emissions}
\label{appsub:ColDen_v_Emission}

\begin{figure*}
    \includegraphics[width=0.80\textwidth]{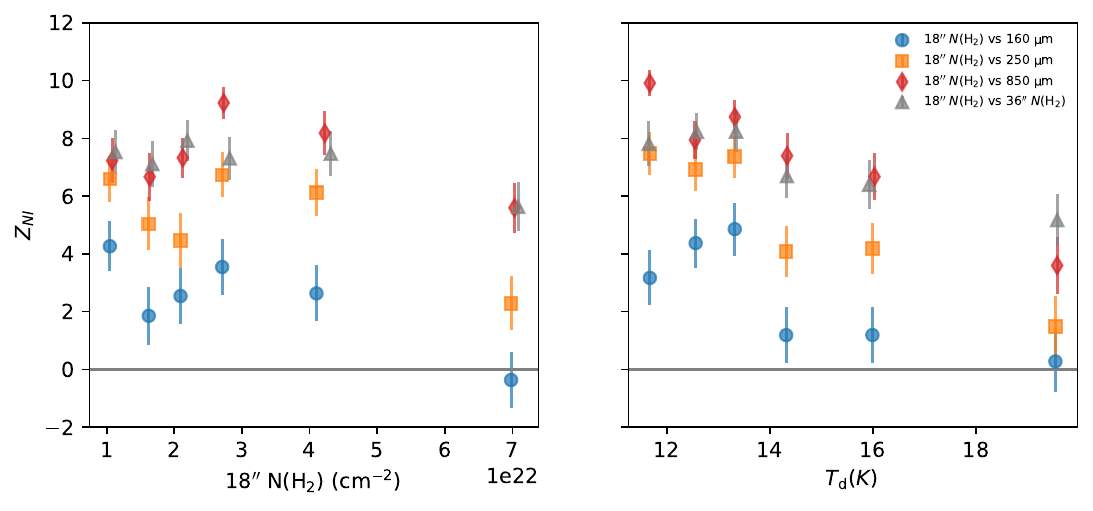}
    \caption{The $Z_{NI}$ values calculated from equal-size bins of $N$(H$_2$) and $T_\mathrm{d}$ plotted against their median bin values in the left and right panels, respectively. The size of each bin is 70 pixels, corresponding to about 14-percentile sampling bin widths. The legend shows the gradients used to calculate the $Z_{NI}$ values.}
    \label{fig:Z_NI_vs_props}
\end{figure*}

\begin{figure*}
    \includegraphics[width=0.80\textwidth]{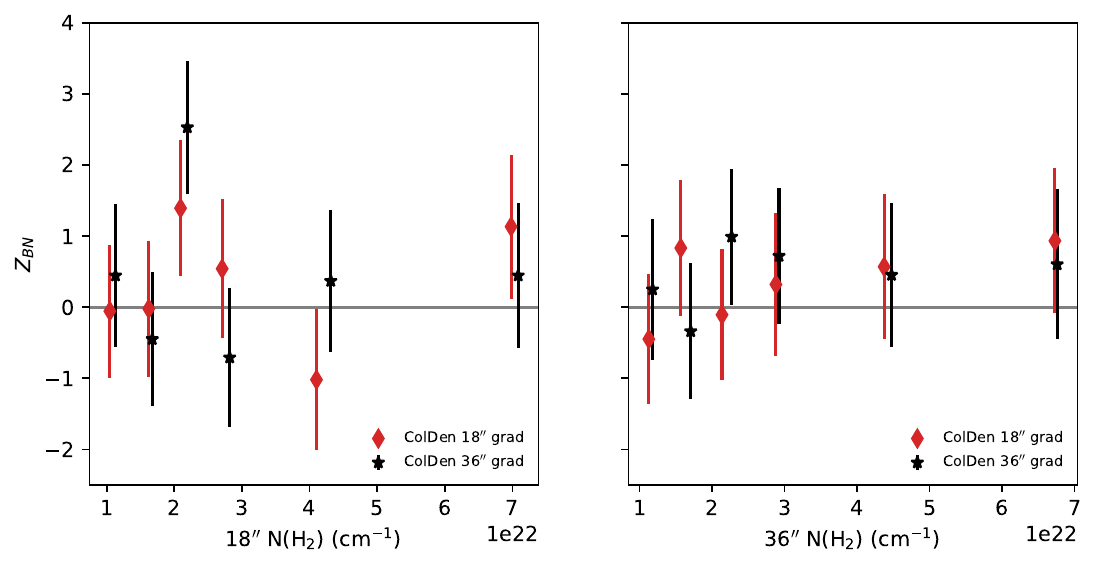}
    \caption{The $Z_{BN}$ values calculated from equal-size bins of $18''$-resolution $N$(H$_2$) and $36''$-resolution $N$(H$_2$) plotted against their median bin values in the left and right panels, respectively. The size of each bin is 70 pixels, corresponding to about 14-percentile sampling bin widths. The $Z_{BN}$ are calculated from gradients of the $18''$- and $36''$-resolution $N$(H$_2$) values, labelled in the legend.}
    \label{fig:Z_BN_reso}
\end{figure*}

To investigate the relationship between the $ 18''$-resolution \textit{Herschel}-derived column densities and the dust emission maps, we calculate the PRS between their respective gradients (i.e., $Z_{NI}$). Figure \ref{fig:Z_NI_vs_props} shows $Z_{NI}$ calculated for 160 \textmu m, 250 \textmu m, and 850 \textmu m maps plotted as a function of $18''$-resolution $N$(H$_2$) and $T_d$ in the left and right panels, respectively. We also applied the same analysis comparing the gradients between the $18''$ and $36''$ resolution $N$(H$_2$) maps and plotted over the same panels.  Overall, we find parallel alignments between the respective ROs, indicating that the column density and dust emission maps broadly trace similar structures. The emission alignments with column density, however, become less parallel as the wavelength decreases. This trend is consistent with the expectation that shorter wavelength emissions preferentially trace warmer structures along the lines of sight than purely column-density-weighted structures. We also see a general anti-correlation between $Z_{NI}$ and $T_d$, indicating that dust emission structures indeed depart more from their column density counterparts in warmer environments.

\subsection{Resolution Dependency}
\label{appsub:ColDen_Resolution}

Figure \ref{fig:Z_NI_vs_props} also shows that the $14''$-resolution 850 \textmu m emission gradients and $36''$-resolution $N$(H$_2$) gradients have a similar degree of alignment with the $18''$-resolution $N$(H$_2$) gradients. This result illustrates that: 1) 850 \textmu m emission is indeed a good proxy of column density to first order and 2) our alignment results are not particularly sensitive to the resolution difference between the $18-$ and $ 36'$-resolution $N$(H$_2$) maps.

To further investigate the effect of resolution on the column density gradients, Figure \ref{fig:Z_BN_reso} shows the PRS between the magnetic field and the gradients of each column density map ($18''$- and $36''$- resolution) plotted relative to equally-sized bins in column density. Both column density maps show similar $Z_{BN}$ values for each column density bin, indicating that $Z_{BN}$ is not particularly sensitive to spatial resolution. The lack of alignment (i.e., $|Z_{BN}| \lesssim 1 \sigma$) in any of the $36''$- resolution bins (left panel) compared to their $18''$- resolution counterpart (right panel), indicates that the $36''$- resolution $N$(H$_2$) is spatially too coarse to resolve $N$(H$_2$) ranges with particularly interesting alignments, such as that seen in $18''$- resolution's $2.5 \times 10^{22}$ cm$^{-2}$ bin. This result reinforces our choice to carry out our analyses using the $18''$-resoltuion $N$(H$_2$) data.


\bsp	
\label{lastpage}
\end{document}